\documentclass{article}
\usepackage{amssymb}

\input{tcilatex}
\begin{document}

\begin{center}
\bigskip

CAN\ QUANTUM THEORY\ BE UNDERPINNED BY A NON-LOCAL HIDDEN\ VARIABLE\ THEORY ?

{\LARGE \bigskip }

B J Dalton{\LARGE \bigskip }

Centre for Quantum Science and Technology Theory, Swinburne University of
Technology, Melbourne, Victoria 3122, Australia\textit{\bigskip }

\textit{\bigskip }

\pagebreak\ 
\end{center}

\subsection{Abstract}

In this paper we consider the description by a general Bell-type non-local
hidden variable theory of bipartite quantum states with two observables per
sub-system. We derive Bell inequalities of the
Collins-Gisin.-Liden-Massar-Popescu type which involve combinations of the
probabilities of related outcomes for measurements for the four pairs of
sub-system observables. It is shown that the corresponding quantum theory
expressions violate the Bell inequalities in the case of the maximally
entangled state of the bipartitite system. The CHSH Bell inequality is also
derived from this general CGLMP Bell-type non-local hidden variable theory.
This shows that quantum theory can not be underpinned by a Bell-type
non-local hidden variable theory. So as a general Bell-type local hidden
variable theory has already been shown to conflict with quantum theory, it
follows that quantum theory can not be understood in terms of any CGLMP
Bell-type hidden variable theory - local or non-local. \bigskip

\pagebreak

\section{INTRODUCTION}

This paper deals with the issue of whether quantum theory needs to be
completed or underpinned by a \emph{non-local hidden variable theory }%
(NLHVT), an issue that originates from the 1930's papers of \emph{Einstein} 
\cite{Einstein35a}{\small \ }and \emph{Schrodinger} (\cite{Schrodinger35a}, 
\cite{Schrodinger35b}{\small ) }and the 1960's paper of \emph{Bell }{\small (%
}\cite{Bell64a}{\small )}. The key word is $\emph{non-local,}$as it is now
widely accepted that quantum theory is \emph{not} underpinned by a \emph{%
local }hidden variable theory (LHVT). \smallskip

\subsection{Quantum Paradoxes}

The 1930's papers identified two paradoxical features arizing from the \emph{%
Copenhagen interpretation} of \emph{quantum} \emph{theory} (QTHY),\ even for
the simplest case of \emph{bipartite} systems. The first was \textit{%
macroscopic entanglement} -- such as in the Schrodinger cat experiment,
wherein states could exist in \emph{bipartite} systems in which a sub-system
could \emph{simultaneously} occupy two \emph{macroscopically distinct}
states. Here the system state was an \emph{entangled} state of a cat and a
radioactive atom, a quantum superposition state involving \textit{both} the
cat being dead\textit{\ and} the cat being alive. This conflicted with
common sense - the cat should be \textit{either} dead \textit{o}r alive. The
second was \emph{EPR steering} - measuring the \emph{value} for an \emph{%
observable} chosen for \emph{one} sub-system $A$ sometimes could \emph{%
instantaneously} affect the outcome for measuring the \emph{value} of an 
\emph{observable} in a \emph{second} sub-system $B$ that was \emph{spatially
well-separated}. This contradicted \textit{relativity} and for a system
involving two particles in a possible state with a well-defined position
difference (and a well-defined momentum sum), also led to a violation of the 
\textit{Heisenberg Uncertainty Principle} in terms of evidently being able
to assign precise values to both\textbf{\ }the position and momentum of one
of the two particles. So maybe the particle really does have both a definite
position and momentum - and the cat is either dead or alive, whilst the
probabilistic feature of quantum theory is just similar to\ that in
classical statistical mechanics ?\smallskip

\subsection{Quantum Theory Incompleteness ?}

$\smallskip $Thus arose \emph{Einstein's question}: \cite{Einstein35a}. Does
quantum theory (though agreeing with \emph{experiment}) require \emph{%
completion }via an \emph{underlying theory} that determines the \emph{actual}
\emph{values} of \textit{physical quantities} (realism)\textbf{\ }- such as
the position and momentum (for the particle) or being alive or being dead
(for the cat), even if these are hidden within the \emph{probability
distribution} ?{\small \ \ }Since the work of Bohm \cite{Bohm52a} and Bell 
\cite{Bell64a} such underlying (or underpinning)\ theories are now generally
referred to as \textit{hidden variable theories}.{\Large (}HVT\emph{\ }%
).\smallskip\ 

\subsection{Local and Non-Local Hidden Variable Theory}

In response, Bell introduced \cite{Bell64a} {\small \ }a so-called \emph{%
local hidden variable theory} (\emph{LHVT}) for describing the quantum
states of composite systems{\small . }These states are referred to as \emph{%
Bell local states}. In the original \emph{deterministi}c version the \emph{%
hidden variables} (HV)\ determine the \emph{actual} outcome for each \emph{%
separate sub-system}, when any of its observables are measured. \ In the 
\emph{general} \emph{version} of LHVT the HV essentially act like \emph{%
classical} \emph{phase space} variables, and determine the \emph{%
probabilities of outcomes} for each \emph{sub-system} \emph{separately} when
any of its observables are measured \emph{(the locality feature)}. A key
idea is that the HV only depend on \emph{state preparation process }$-$ not
on the subsequent \emph{choice }of what observables will be \emph{measured}.

LHVT provides predictions for \emph{inequalities} involving results based on 
\emph{mean values} or the \emph{probabilities} of joint measurement outcomes
of \emph{pairs} of observables from the \emph{two} sub-systems - generically
referred to as \emph{Bell} \emph{inequalities}. We can then \emph{test}
whether the \emph{inequality predictions} based on \emph{LHVT} are
consistent both with \emph{quantum theory} predictions and with \emph{%
experiment}. As described below, \emph{some} Bell inequality violations show 
\emph{LHVT} fails. So does this mean that quantum theory should be
underpinned or completed via a\emph{\ Bell-type non-local} \emph{HVT }%
(NLHVT) ?

In this paper a general \emph{non-local} hidden variable theory will be
presented, and Bell inequalities obtained based on this NLHVT.We will show
that the famous CHSH\ \cite{Clauser69a} Bell inequality can be established
from NLHVT as well as from LHVT. We then consider Bell inequalities\textbf{\ 
}of the Collins, Gisin, Linden, Massar Popescu (CGLMP)\ type (see \cite%
{Collins02a}, \cite{Dalton21a}). It will then be shown that in a \emph{%
maximally entangled} state in a bipartite system there are\textbf{\ }clear
violations of these Bell inequalities, thus showing that it is \emph{not
posible} to underpin or complete quantum theory via this general Bell-type
non-local hidden variable theory. As well as being an important \emph{%
quantum foundations }issue, there are also implications also for \emph{%
quantum technology, since systems for }secure communications, quantum
information processing, etc depend on there being different outcomes
predicted for quantum theory and for hidden variable theory.\medskip\ 

\subsection{Bohmian Mechanics}

\label{SubSectionBohmianMechanics}

It should be noted that previous to the work of Bell \cite{Bell64a}, a \emph{%
different} kind of non-local hidden variable theory was introduced by Bohm 
\cite{Bohm52a} in the 1950's, based on earlier work by de Broglie\ (see \cite%
{Durr20a} for a recent account). The approach involves\emph{\ deterministic} 
\emph{particle trajectories} dependent on a so-called \emph{quantum\
potential} determined (in the non-relativistic version) from the
multi-particle \emph{Schrodinger wave function }for the system\emph{. }The 
\emph{particle positions} act like \emph{hidden variables} and non-locality
refers to the feature that the \emph{velocity} of any one particle depends
on the positions of \textit{all }the other particles, no matter where their
relative locations may be. This would violate \emph{special relativity}, but
as the related version of quantum theory is also non-relativistic this
outcome is merely being consistent. As the derivation of the equations for
Bohmian mechanics (BM) invokes the existence of the wave-function, assumes
the correctness of the \emph{time-dependent Schrodinger equation }(TDSE)%
\emph{\ }and that the \emph{modulus squared} of the wave function is the 
\emph{position probability density }for the particles, it may be concluded
that Bohmian mechanics is based on just the same equations as quantum theory
but with some \emph{extra assumptions} and \emph{ontological features} (the
deterministic trajectories) \emph{added}. The particle trajectory feature is
based on deriving a continuity equation for the position probabilty density
and identifying the\emph{\ gradient} of the \emph{phase} $S$ of the
wave-function (divided by the partticle mass) with the \textit{velocity} for
the deterministic particle trajectory\textbf{\ }- whose existence is now
incorporated into the theory. This identification is based on a continuity
equation for the position probability derived from equations for the phase
and amplitude of the wave-function obtained from the TDSE. So the addition
of deterministic trajectories complies with Einstein's suggestion of \emph{%
completing} quantum theory by adding\ new features - the deterministic
trajectories. Whether the particle positions should be regarded as "hidden
variables" is a moot point. Unlike the hidden variables $\lambda $
introduced by Bell \cite{Bell64a} (see below), the particle positions could
in principle be observed, and they obey well defined Hamilton-Jacobi
equations involving the quantum potential - in contrast to the unspecified
equations for the $P(\lambda |c)$ and $P(\alpha |\Omega _{A},\lambda )$ in
Bell's approach. BM is rather like classical \ physics and uses a more
familiar mathematics based on calculus rather than the mathematics of linear
vector spaces and Hermitian \ operators. Many of the characteristic results
in standard quantum theory can also be shown in BM. For example the
component $L_{z}$ of the orbital \ angular momentum of a particle is $%
xp_{y}-yp_{x}$, which becomes $x(\partial S/\partial y)-y(\partial
S/\partial x)$ $=(\partial S/\partial \phi )$ from the particle trajectory
assumption in which $p_{y}=mv_{y}$ etc.. As the \emph{stationary} solution
of the Schrodinger equation for a spherical potential involves a phase $%
S=\beta -Et+\hslash m\phi $ ($m$ an integer) it follows that $L_{z}$ $\ $has
a \emph{quantised} value $m\hslash $. Other quantum theory features such as
the \emph{symmetry} or \emph{anti-symmetry} of the wave-function for
identical particles are just \emph{assumed} true in Bohmian mechanics. \emph{%
Relativistic versions} of Bohmian mechanics are still being developed, as
are \emph{quantum field theory} versions. As far as is known, Bohmian
mechanics does not predict any \emph{new} physical results, but it has been
applied to the well-known two-slit experiment for single particles \cite%
{Hiley79a} - where it is shown that the particle trajectories are affected
by the quantum potential (that reflects the interference effects on the
wave-function due to the two slits), and this then leads to the periodic
particle position probability distribution on the observation screen.
Whether Bohmian mechanics can describe bipartite systems for \emph{all}
pairs of non-commuting observables in each sub-system is an issue that is
not examined in the present paper. However, interest in Bohmian mechanics
and its extensions still continues \cite{Hall18a}. \medskip

\subsection{Other Proposed Non Local HVT}

\label{SubSectionOtherNonLocalHVT}

Another approach to completing quantum theory is the\textbf{\ }\emph{Many
Worlds Interpretation (MWI)\ }\cite{Everett57a}. Essentially the\textbf{\ }%
\emph{probabilistic feature }of quantum theory - where in general each
possible outcome of measuring an observable only occurs with a certain
probability - is replaced with a\textbf{\ }\emph{deterministic picture} in
which all the different outcomes actually occur, but now in dfferent
universes. The MWI has recently been discussed by Vaidman \cite{Vaidman21a}.
Leggett \cite{Leggett03a}, introduced a so-called\textbf{\ }\emph{%
Crypto-Nonlocal (CN)} hidden variable theory.In the context of the present
paper, the LHVT\ expression (\ref{Eq.LHVTJointProb}) $P(\alpha |\Omega
_{A},\lambda )P(\beta |\Omega _{B},\lambda )$\ for the bipartite system case
would be replaced in the CN\ theory by $P(\alpha |\Omega _{A},\Omega
_{B},\lambda )P(\beta |\Omega _{A},\Omega _{B},\lambda )$, so that although
the\textbf{\ }\emph{outcome} for measuring either sub-system observable does
not depend on the outcome for measuring the other sub-system observable, it
does depend on the\textbf{\ }\emph{choice}\textbf{\ }of the other
observable. It is not clear whether this would produce different Bell
inequalities from the NLHVT replacement $P(\alpha ,\beta |\Omega _{A},\Omega
_{B},\lambda )$. CN theory has been found to violate quantum theory and
experiment \cite{Branciard07a}\textbf{. \medskip }\ 

\subsection{Plan of Paper}

First however, we need to outline the key equations describing \emph{both}
LHVT\ \emph{and} a form of NLHVT in the simplest case of bipartite systems.
Here the basic probability involves the outcomes for\textbf{\ }\emph{one}%
\textbf{\ }observable in\textbf{\ }\emph{each}\textbf{\ }sub-system. We then
discuss the well-known CHSH\ example of a Bell inequality based on LHVT\
that is inconsistent with quantum theory. We also outline the idea behind
LHVT. For reasons of completeness we next\textbf{\ }point out how Bell
inequalities are also relevant to discussing quantum entanglement and EPR
steering based on \ a classification of Bell local states into three
categories involving so-called \emph{local hidden quantum states} \cite%
{Wiseman07a}. The material outlining this is in Appendix A. We then
introduce the so-called CGLMP\ Bell inequalities \cite{Collins02a}. Here the
basic probability involves the outcomes for two observables in\textbf{\ }%
\emph{each}\textbf{\ }sub-system, resulting in more general LHVT andNLHVT\
forms. We then show a novel result - namely that the CHSH Bell inequality can%
\textbf{\ }\emph{also}\textbf{\ }be proved for this more general NLHVT. The
violation of the CHSH Bell inequality for this NLHVT\ approach proves that
quantum theory cannot be underpinned by this type of NLHVT. The CGLMP
inequalities themselves are applied in a bipartite system where each
sub-system has\textbf{\ }\emph{two}\textbf{\ }observables, each with the%
\textbf{\ }\emph{same} number $d$ of outcomes) and set out the relevant
equations for the inequalities - involving the quantity $S$, and introduce
the observables and the test quantum state, both for the LHVT and NLHVT\
cases. The quantity $S$ involves the\textbf{\ }\emph{probabilities}\textbf{%
\emph{\ }}of \emph{all} the outcomes for \emph{pairs of observables}\textbf{%
\ }in the two sub-systems, where the outcomes are\textbf{\emph{\ }}\emph{%
shifted} by fixed amounts.\textbf{\ }The next section covers the basic \emph{%
matrix equation method} for determining \emph{sufficiency}\textbf{\ }%
requirements for possible Bell inequalities of the form $S\leq 4,S\leq
3,S\leq 2,S\leq 1$ in\emph{\ four} cases for the NLHVT situation \ Based on
a consideration of\emph{\ all} the \emph{sub-cases}\textbf{, }sufficiency
conditions\textbf{\emph{\ }for} Bell \textbf{tests} in each of these\textbf{%
\ }four cases are set out, with the detailed derivations being presented in
Appendix B. The \emph{number} of sub-cases is $1,4,6,4$\ for $S\leq 4,S\leq
3,S\leq 2,S\leq 1$\ respectively. The conditions are in the form of\textbf{\ 
}\emph{inequalities} involving the \emph{outcome shifts}, and for the cases
of $S\leq 2,S\leq 1$\ , there are\textbf{\ }\emph{more than one}\textbf{\ }%
such inequality for \emph{each} of the sub-cases that are involved. At least
one inequality must be satisfied for every sub-case. The last part of this
section then sets out the quantum theory expressions for $S$. A comparison
of the quantum theory numerical results for $S$ with the NLHVT\ results then
follows. We again show that the CGLMP Bell inequality based on NLHVT is
violated in quantum theory, reinforcing the previous conclusion based on
CHSH. The paper ends with a Summary and Conclusions. \medskip\ 

\section{QTHY \& HVT BASIC EQUATIONS}

We consider the case of a bipartite system in which the two sub-systems may
be spatialy separated. At \ present we only consider \emph{one} observable
per sub-system- the \ case of \emph{two} will be considered later.

\subsection{QTHY \& HVT\ Probabilities - Bipartite Systems}

The \emph{QTHY} and \emph{LHVT} \emph{joint probabilities} $P(\alpha ,\beta
|\Omega _{A},\Omega _{B})$ for measurements of \emph{observables} $\Omega
_{A}$, $\Omega _{B}$ for sub-systems $A$, $B$ with \emph{outcomes} $\alpha $%
, $\beta $ are: 
\begin{eqnarray}
P(\alpha ,\beta |\Omega _{A},\Omega _{B},\rho )_{Q} &=&Tr((\widehat{\Pi }%
_{\alpha }^{A}\otimes \widehat{\Pi }_{\beta }^{B})\widehat{\rho })
\label{Eq.QThyJointProb} \\
P(\alpha ,\beta |\Omega _{A},\Omega _{B},c)_{LHVT} &=&\dsum\limits_{\lambda
}P(\lambda |c)P(\alpha |\Omega _{A},\lambda )P(\beta |\Omega _{B},\lambda )
\label{Eq.LHVTJointProb}
\end{eqnarray}

For QTHY $\widehat{\Pi }_{\alpha }^{A}$ is the \emph{projector} onto the
space of \emph{eigenvectors} for \emph{operator} $\widehat{\Omega }_{A}$
with \emph{eigenvalue} $\alpha $. $\widehat{\rho }$ is \emph{density opr}
specifying the \emph{preparation} \emph{process} $c$ of the\textbf{\ }%
quantum state. For LHVT, \emph{hidden variables} $\lambda $ determined with
probability $P(\lambda |c)$ for preparation process $c$. The projectors
satisfy $\sum_{\alpha }\widehat{\Pi }_{\alpha }^{A}=\widehat{1}^{A}$ and $%
\sum_{\beta }\widehat{\Pi }_{\beta }^{B}=\widehat{1}^{B}.$

In the \emph{probabilistic} form of LHVT the \emph{hidden variables} $%
\lambda $ determine each \emph{local sub-system} measurement \emph{%
probabilities\thinspace\ }$P(\alpha |\Omega _{A},\lambda )$, $P(\beta
|\Omega _{B},\lambda )$ for \emph{classical observables} $\Omega _{A}$, $%
\Omega _{B}$ - which \ are combined using \emph{classical probability theory}
to determine joint measurement probabilities. Since the probabilities for%
\emph{\ all }outcomes must add up to \emph{unity} we have $\sum_{\alpha
}P(\alpha |\Omega _{A},\lambda )=1$ and $\sum_{\beta }P(\beta |\Omega
_{B},\lambda )=1$. The origin of this expression for the LHVT probability
expression involving the product of \emph{separate}\textbf{\ }sub-system
probabilities is set out below in Section \ref{SubsysOriginLHVTExpression}.
This form for the LHVT\ probability is widely used - apart from \cite%
{Bell64a}, see for example Refs. \cite{Wiseman07a}, \cite{Reid09a}, \cite%
{Durr20a}.

Both the QTHY and the LHVT probability satisfy the\textbf{\ }\emph{%
no-signaling}\textbf{\ }condition. We have \textbf{%
\begin{eqnarray}
\sum_{\alpha }P(\alpha ,\beta |\Omega _{A},\Omega _{B},c)_{Q} &=&Tr((%
\widehat{1}^{A}\otimes \widehat{\Pi }_{\beta }^{B})\widehat{\rho }%
)=\sum_{\alpha ^{\#}}P(\alpha ^{\#},\beta |\Omega _{A}^{\#},\Omega
_{B},c)_{Q}  \nonumber \\
\sum_{\beta }P(\alpha ,\beta |\Omega _{A},\Omega _{B},c)_{Q} &=&Tr((\widehat{%
\Pi }_{\alpha }^{A}\otimes \widehat{1}^{B})\widehat{\rho })=\sum_{\beta
^{\#}}P(\alpha ,\beta ^{\#}|\Omega _{A},\Omega _{B}^{\#},c)_{Q}  \nonumber \\
\sum_{\alpha }P(\alpha ,\beta |\Omega _{A},\Omega _{B},c)_{LHVT}
&=&\dsum\limits_{\lambda }P(\lambda |c)P(\beta |\Omega _{B},\lambda
)=\sum_{\alpha ^{\#}}P(\alpha ^{\#},\beta |\Omega _{A}^{\#},\Omega
_{B},c)_{LHVT}  \nonumber \\
\sum_{\beta }P(\alpha ,\beta |\Omega _{A},\Omega _{B},c)_{LHVT}
&=&\dsum\limits_{\lambda }P(\lambda |c)P(\alpha |\Omega _{A},\lambda
)=\sum_{\beta ^{\#}}P(\alpha ,\beta ^{\#}|\Omega _{A},\Omega
_{B}^{\#},c)_{LHVT}  \nonumber \\
&&  \label{Eq.NoSignalCondn}
\end{eqnarray}%
}where two different choices of sub-system observables $\Omega
_{A}^{\#},\Omega _{B}^{\#}$ and outcomes $\alpha ^{\#},\beta ^{\#}$ are
considered. Thus the outcome for sub-system's observable is\textbf{\ }\emph{%
unaffected}\textbf{\ }by the\textbf{\ }\emph{outcome }or\textbf{\ }\emph{%
choice}\textbf{\ }of the other sub-system's observable.

If the LHVT form for the probability $P(\alpha ,\beta |\Omega _{A},\Omega
_{B},c)_{LHVT}$ does not apply, then what form would apply for NLHVT ? As
explained in Sect. \ref{SubsysOriginLHVTExpression} the locality condition
results in $P(\alpha ,\beta |\Omega _{A},\Omega _{B},c)_{LHVT}$ involving
the product of separate sub-system outcome probabilities as in Eq (\ref%
{Eq.LHVTJointProb}). Hence a general NLHVT\ probability will be one which \
does not involve separate sub-system probabilities. Thus, in the \emph{%
probabilistic} form of this general type of\textbf{\ }NLHVT, the \emph{%
hidden variables} $\lambda $ should determine the \emph{combined sub-system}
measurement \emph{probabilities\thinspace\ }$P(\alpha ,\beta |\Omega
_{A},\Omega _{B},\lambda )$, for \emph{both}\textbf{\ }\emph{classical
observables} $\Omega _{A}$, $\Omega _{B}$ \ In this general\textbf{\ }%
non--local hidden variable theory there is \emph{no separate} (or local) HVT
probability for each sub-system.%
\begin{equation}
P(\alpha ,\beta |\Omega _{A},\Omega _{B},c)_{NLHVT}=\dsum\limits_{\lambda
}P(\lambda |c)P(\alpha ,\beta |\Omega _{A},\Omega _{B},\lambda )
\label{Eq.NLHVTJointProb}
\end{equation}%
Again, the sum of the probabilities for all outcomes must be unity,so $%
\sum_{\alpha ,\beta }P(\alpha ,\beta |\Omega _{A},\Omega _{B},\lambda )=1$.
This form for the NLHVT proability is the general form that would be
involved when the local form in (\ref{Eq.LHVTJointProb}) is not applicable.
Other more specific forms of NLHVT\ (such as Bohmian Mechanics) are briefly
discussed in Sects. \ref{SubSectionBohmianMechanics} and \ref%
{SubSectionOtherNonLocalHVT}. There is no obvious reason why the\textbf{\ }%
\emph{no-signaling }condition would apply in NLHVT.\textbf{\ }\medskip\ 

\subsection{Joint Measurement Mean Values}

QTHY and LHVT expressions for the \emph{mean values} of joint measurement
outcomes for $\Omega _{A},\Omega _{B}$ are: 
\begin{eqnarray}
\left\langle \Omega _{A}\otimes \Omega _{B}\right\rangle _{Q} &=&Tr(\widehat{%
\Omega }_{A}\otimes \widehat{\Omega }_{B})\widehat{\rho } \\
\left\langle \Omega _{A}\otimes \Omega _{B}\right\rangle _{LHVT}
&=&\dsum\limits_{\lambda }P(\lambda |c)\left\langle \Omega _{A}(\lambda
)\right\rangle \left\langle \Omega _{B}(\lambda )\right\rangle
\end{eqnarray}%
where $\left\langle \Omega _{A}(\lambda )\right\rangle =\sum_{\alpha }\alpha
P(\alpha |\Omega _{A},\lambda )$ is LHVT mean value for measurement of $%
\Omega _{A}$ when hidden variables are $\lambda $. Similarly for\textbf{\ }$%
\left\langle \Omega _{B}(\lambda )\right\rangle $.

The NLHVT expression for the \emph{mean values} of joint measurement
outcomes for $\Omega _{A},\Omega _{B}$ is%
\begin{equation}
\left\langle \Omega _{A}\otimes \Omega _{B}\right\rangle
_{NLHVT}=\dsum\limits_{\lambda }P(\lambda |c)\left\langle (\Omega
_{A}\otimes \Omega _{B})(\lambda )\right\rangle
\label{Eq.MeanJointMeastNLHVT}
\end{equation}%
where $\left\langle (\Omega _{A}\otimes \Omega _{B})(\lambda )\right\rangle
=\sum_{\alpha ,\beta }\alpha \beta \,P(\alpha ,\beta |\Omega _{A},\Omega
_{B},\lambda )$ is the NLHVT mean value for measurement of $\Omega _{A}$ and 
$\Omega _{B}$ when hidden variables are $\lambda $.\smallskip

\subsection{CHSH\ Bell Inequality Test}

\label{SubSysCHSHTest}

The well-known \emph{CHSH }Bell inequality (\cite{Clauser69a})\ for
bipartite system where outcomes $\alpha ,\beta $ are restricted to $(+1,-1)$
and involving \emph{two measurement choices} for each sub-system is 
\begin{eqnarray}
|S| &=&|\left\langle \Omega _{A1}\otimes \Omega _{B1}\right\rangle
_{LHVT}+\left\langle \Omega _{A1}\otimes \Omega _{B2}\right\rangle
_{LHVT}+\left\langle \Omega _{A2}\otimes \Omega _{B1}\right\rangle
_{LHVT}-\left\langle \Omega _{A2}\otimes \Omega _{B2}\right\rangle
_{LHVT}\,|\,\leq 2  \nonumber \\
&&  \label{Eq.CHSHBellIneq}
\end{eqnarray}%
Note the minus sign.\textbf{\ }This is violated for the\textbf{\ }\emph{%
singlet state} of two spin $1/2$ systems (spin operators are in units $\hbar
/2$).%
\begin{equation}
\left\lfloor \Psi \right\rangle =\left( \left\vert +1\right\rangle _{S_{\mu
}^{A}}\left\vert -1\right\rangle _{S_{\mu }^{B}}-\left\vert -1\right\rangle
_{S_{\mu }^{A}}\left\vert +1\right\rangle _{S_{\mu }^{B}}\right) /\sqrt{2}%
\qquad \mu =x,y,or\;z
\end{equation}%
for Pauli spin operators $\Omega _{A1}=\sigma _{z}^{A}$, $\Omega
_{A2}=\sigma _{x}^{A}$, $\Omega _{B1}=-(\sigma _{x}^{B}+\sigma _{z}^{B})/%
\sqrt{2}$, $\Omega _{B2}=+(\sigma _{x}^{B}-\sigma _{z}^{B})/\sqrt{2}$, for
which we find $S=2\sqrt{2}$ from quantum theory. This is based on the
quantum theory mean value for the product of Pauli spin operator components
along directions specified by unit vectors $\overrightarrow{a}$ and $%
\overrightarrow{b}$ given by $\left\langle \overrightarrow{\sigma }^{A}\cdot 
\overrightarrow{a}\otimes \overrightarrow{\sigma }^{B}\cdot \overrightarrow{b%
}\right\rangle _{Q}=-\overrightarrow{a}\cdot \overrightarrow{b}$ for the
singlet state.

Note that the proof of the CHSH Bell inequality is based on the LHVT form (%
\ref{Eq.LHVTJointProb}) for the joint probability. It\textbf{\ }\emph{cannot}%
\textbf{\ }be proved from the NLHVT form (\ref{Eq.NLHVTJointProb}), though
as we will see in Sect \ref{SubsectProof of CHSH Bell Inequality}, the CHSH
Bell inequality\textbf{\ }\emph{can}\textbf{\ }be proved based on the\textbf{%
\ }\emph{two}\textbf{\ }observable per sub-system NLHVT\ joint probability
form (\ref{Eq.NLHVTCGLMP}).

Thus in some Bell inequalities (such as the \emph{CHSH inequalities} \cite%
{Clauser69a}) quantum states were found in \emph{microscopic} bipartite
systems where \emph{failure} of Bell LHVT was\emph{\ predicted}, and \emph{%
experiments} {\small (}\cite{Aspect82a}, \cite{Zeilinger97a}{\small ) }\emph{%
agreed} with quantum theory. This was sufficient to rule out \emph{LHVT}
accounting for quantum theory \emph{in general}. \smallskip

\subsection{Origin of LHVT Expression}

\label{SubsysOriginLHVTExpression}

The origin of the LHVT joint probability expression (\ref{Eq.LHVTJointProb})
is explained in Ref \cite{Durr20a}. Consider two \textit{events} - $A$ being
the measurement of observable $\Omega _{A}$ with outcome $\alpha $ and $B$
being the measurement of observable $\Omega _{B}$ with outcome $\beta $ ,
where in both cases when the hidden variables are $\lambda $. These events
have a joint probability $P(B,A)$ for \emph{both} occuring, whilst the
probability of the \emph{separate }events are $P(A)$, $P(B)$. Consider a
situation (such as may occur when the events occur for separate widely
separated sub-systems) where \ the \emph{conditional probability} $P(B|A)$
for event $B$ occuring \emph{given} that event $A$ \emph{occurs} has the
same probability \textit{as if }event $A$ \ does \emph{not} occur. This is
the situation of \textit{locality}. Here the conditional probability $P(B|A)$
is independent of event $A$ and thus will not depend on the probability $%
P(A) $ of event $A$ occuring. 
\begin{eqnarray}
P(B|A) &=&P(B) \\
P(B,A) &=&P(B|A)P(A) \\
&=&P(B)\times P(A)=P(A)\times P(B)=P(A,B)
\end{eqnarray}%
after applying Bayes' theorem\textbf{. }Thus the probability $P(B,A)=P(A,B)$
of both events occuring is equal to the product of the probabilities of each
separate event occuring, and\textbf{\ }we obtain the LHVT expression (\ref%
{Eq.LHVTJointProb}) 
\begin{eqnarray}
P(\alpha ,\beta |\Omega _{A},\Omega _{B},\lambda )_{LHVT} &=&P(\alpha
|\Omega _{A},\lambda )P(\beta |\Omega _{B},\lambda )  \label{Eq.LHVTFactn} \\
P(\alpha ,\beta |\Omega _{A},\Omega _{B},c)_{LHVT} &=&\dsum\limits_{\lambda
}P(\lambda |c)P(\alpha |\Omega _{A},\lambda )P(\beta |\Omega _{B},\lambda )
\end{eqnarray}%
after averaging over all hidden \ variables $\lambda $- which occur with
probability $P(\lambda |c)$. for preparation process $c$.

In the case where the hidden variable theory is non-local\textbf{,} the
factorisation in Eq,.(\ref{Eq.LHVTFactn}) does \emph{not} apply (see Ref 
\cite{Wiseman07a}) \ and $P(\alpha |\Omega _{A},\lambda )P(\beta |\Omega
_{B},\lambda )$ would be replaced by $P(\alpha ,\beta |\Omega _{A},\Omega
_{B},\lambda )$, as in Eq (\ref{Eq.NLHVTJointProb}).\smallskip

\subsection{Quantum Entanglement and EPR Steering ?}

\emph{Bell locality} violation also implies that \emph{EPR steering} and 
\emph{quantum entanglement} both occur. However quantum entanglement or EPR
steering can \emph{also} occur for \emph{some} Bell local states (\cite%
{Werner89a}, \cite{Wiseman07a}, \cite{Jones07a}, \cite{Cavalcanti09a}, \cite%
{Dalton20a}). Although this paper is focused on tests for \emph{Bell locality%
} violation, Bell tests for \emph{quantum entanglement} or \emph{EPR
steering }are also\emph{\ }important \emph{(}\cite{Dalton20a}{\small ). }%
After all, issues regarding these effects were part of the motivation for
the search for a theory to underpin quantum theory.\textbf{\ }For
completeness, a brief outline of how LHVT Bell states can be divided into 
\emph{three} categories with different featues for Quantum Entanglement and
EPR Steering is set out in Appendix A. The violation of \emph{other} Bell
inequalities for LHVT based on this categorisation\textbf{\ }involving \emph{%
local hidden quantum states }(\emph{LHS}) \cite{Wiseman07a}{\small \ }show
whether or not \emph{Quantum entanglement }or\emph{\ EPR steering} is
occuring.

Bell inequalities also exist for \emph{multi-partite} systems such as in%
\textbf{\ }GHZ states \cite{Greenberger90a}{\small \ }or for measurements at 
\emph{three} different times as in \ the\textbf{\ } \emph{Leggett-Garg
inequalities }(\cite{Leggett80a}).\emph{\ \ }A recent review dealing with
Bell correlations in \emph{macroscopic} and \emph{mesoscopic} systems is in
Ref.\cite{Teh21a}\emph{\ }

In the present paper we will be focused on the so-called \emph{CGLMP Bell
inequatities} (\cite{Collins02a}, \cite{Dalton21a}) for bipartite systems.
\pagebreak

\section{CGLMP BELL\ INEQUALITIES}

The CGLMP\ Bell inequalities will now be discussed - both for the \emph{\
local} hidden variable theory case and for the \emph{non-local} hidden
variable theory case. Here we consider the case of a bipartite system in
which the two sub-systems may be spatialy separated, but now.we consider 
\emph{two} observables per sub-system.\textbf{\ }\smallskip

\subsection{General CGLMP\ Considerations}

A \emph{bipartite} system with \emph{two} observables per sub-system, based
on a \emph{HVT} joint probability $P(\alpha _{j},\alpha _{k},\beta
_{l},\beta _{m}|\Omega _{A1},\Omega _{A2},\Omega _{B1},\Omega _{B2})\equiv
C(j,k,l,m)$ for all \emph{four} observables $\Omega _{A1},\Omega
_{A2},\Omega _{B1},\Omega _{B2}$ (\cite{Collins02a}, \cite{Dalton21a}) is
considered. To \emph{shorten} the notation we may write $\alpha _{j},\alpha
_{k},\beta _{l},\beta _{m}\equiv j,k,l,m$. In this bipartite case there are
the \emph{same} number $d$ of \emph{outcomes }for each observable\emph{\ }%
listed $0,1,...,d-1$ for each observable listed as\textbf{\ }$\ j,k,l,m$..

CGLMP (\cite{Collins02a}) wrote their paper for a\textbf{\ }\emph{%
deterministic }version of HVT. Here the hidden variables are the\textbf{\ }%
\emph{outcomes}\textbf{\ }$\alpha _{j},\alpha _{k},\beta _{l},\beta _{m}$
themselves. However,they pointed out that their theory also has a\textbf{\ }%
\emph{probabilistic }version,- in which hidden variables $\lambda $ only
determine the probabilities $P(\alpha _{j},\alpha _{k},\beta _{l},\beta
_{m}|\Omega _{A1},\Omega _{A2},\Omega _{B1},\Omega _{B2};\lambda )$ of the
outcomes. for particular $\lambda $. \ In this probabilistic case we would
have\textbf{\ }%
\begin{eqnarray}
P(\alpha _{j},\alpha _{k},\beta _{l},\beta _{m}|\Omega _{A1},\Omega
_{A2},\Omega _{B1},\Omega _{B2}) &=&\dsum\limits_{\lambda }P(\lambda
|c)P(\alpha _{j},\alpha _{k},\beta _{l},\beta _{m}|\Omega _{A1},\Omega
_{A2},\Omega _{B1},\Omega _{B2};\lambda )  \nonumber \\
&&  \label{Eq.NLHVTCGLMP}
\end{eqnarray}%
For each HV choice $\lambda $ the probabilities for all outcomes must add to
unity, so $\sum_{jklm}P(\alpha _{j},\alpha _{k},\beta _{l},\beta _{m}|\Omega
_{A1},\Omega _{A2},\Omega _{B1},\Omega _{B2};\lambda )=1$. This form for $%
P(\alpha _{j},\alpha _{k},\beta _{l},\beta _{m}|\Omega _{A1},\Omega
_{A2},\Omega _{B1},\Omega _{B2})$ applies for\emph{\ both}\textbf{\ }the
LHVT and NLHVT situations, with the (below) LHVT form (\ref{Eq.LocalHVTCGLMP}%
) involving a factorisation of $P(\alpha _{j},\alpha _{k},\beta _{l},\beta
_{m}|\Omega _{A1},\Omega _{A2},\Omega _{B1},\Omega _{B2};\lambda )$.

There is no reason why the HVT cannot be a NLHVT whose basic probability is
given by Eq (\ref{Eq.NLHVTCGLMP}) and which involves \emph{two} observables
per sub-system. After all, this is consistent with the idea of \emph{realism}
in which the hidden variables are the \emph{actual }outcomes $\alpha
_{1},\alpha _{2},\beta _{1},\beta _{2}$ of the observables $\Omega
_{A1},\Omega _{A2},\Omega _{B1},\Omega _{B2}$ in the \emph{deternministic}
version or their\ \emph{probabilities }$P(\alpha _{1},\alpha _{2},\beta
_{1},\beta _{2}|\Omega _{A1},\Omega _{A2},\Omega _{B1},\Omega _{B2};\lambda
) $ in the \emph{probabilistic} version with \ hidden variables $\lambda $.
Such an approach is the basis of the.CGLMP\ paper. In both probabilistic and
deterministic versions the outcomes have a real existence prior to any
measurements, as Einstein wanted in any theory underpinning quantum theory.
Also in hidden variable theory, the observables are \emph{not} non-commuting
Hermitian operators, so it is \emph{legitimate} to have an approach where 
\emph{all }obervables (commuting or non -commuting in quantum theory) have 
\emph{simultaneous} \emph{outcomes}. In Bohmian Mechanics the positions and
momenta of every particle is ascribed a real existence. Of course for
comparison with quantum theory or experiment we must construct expressions
that correspond to quantitities obtainable from quantum theory formulae or
which can \ be measured. However this can be achieved - as for quantities
such as $\left\langle \Omega _{Ai}\otimes \Omega _{Bj}\right\rangle $, which
in \emph{quantum theory} are given by $Tr(\widehat{\Omega }_{Ai}\otimes 
\widehat{\Omega }_{Bj}\;\widehat{\rho })$ or in \emph{experiment} by
repeated measurements on each pair of observables $\Omega _{Ai}$ and $\Omega
_{Bj}$.to determine the mean value of the product of the outcomes.

CGMP also only considered cases where the outcomes for one of the
observables for each sub-system was considered - $\Omega _{A1}$ or $\Omega
_{A2}$ for sub-system $A$ and $\Omega _{B1}$ or $\Omega _{B2}$ for
sub-system $B.$ We restrict ourselves to this situation, and hence the only
HVT\textbf{\ }\emph{marginal probabilities }we consider (in the general
probabilistic version) are\textbf{\ }%
\begin{eqnarray}
P(\alpha _{j},\beta _{l}|\Omega _{A1},\Omega _{B1}) &=&\dsum\limits_{\lambda
}P(\lambda |c)\dsum\limits_{\alpha _{k}\beta _{m}}P(\alpha _{j},\alpha
_{k},\beta _{l},\beta _{m}|\Omega _{A1},\Omega _{A2},\Omega _{B1},\Omega
_{B2};\lambda )  \nonumber \\
P(\alpha _{k},\beta _{l1}|\Omega _{A2},\Omega _{B1})
&=&\dsum\limits_{\lambda }P(\lambda |c)\dsum\limits_{\alpha _{j}\beta
_{m}}P(\alpha _{j},\alpha _{k},\beta _{l},\beta _{m}|\Omega _{A1},\Omega
_{A2},\Omega _{B1},\Omega _{B2};\lambda )  \nonumber \\
P(\alpha _{j},\beta _{m}|\Omega _{A2},\Omega _{B2}) &=&\dsum\limits_{\lambda
}P(\lambda |c)\dsum\limits_{\alpha _{k}\beta _{l}}P(\alpha _{j},\alpha
_{k},\beta _{l},\beta _{m}|\Omega _{A1},\Omega _{A2},\Omega _{B1},\Omega
_{B2};\lambda )  \nonumber \\
P(\alpha _{k},\beta _{m}|\Omega _{A1},\Omega _{B2}) &=&\dsum\limits_{\lambda
}P(\lambda |c)\dsum\limits_{\alpha _{j}\beta _{l}}P(\alpha _{j},\alpha
_{k},\beta _{l},\beta _{m}|\Omega _{A1},\Omega _{A2},\Omega _{B1},\Omega
_{B2};\lambda )  \nonumber \\
&&  \label{Eq.MargProb2}
\end{eqnarray}

So far we have not specified whether the CGLMP \ Bell inequalities we
consider are based on a\textbf{\ }\emph{local}\textbf{\textit{\ }}or a%
\textbf{\ }\emph{non-local}\textbf{\ }version of HVT. \cite{Collins02a}
state that they are considering a local version of HVT. This statement is
rather puzzling as CGLMP do not express $P(\alpha _{1},\alpha _{2},\beta
_{1},\beta _{2}|\Omega _{A1},\Omega _{A2},\Omega _{B1},\Omega _{B2};\lambda
) $\ or $C(j,k,l,m)$ as the\textbf{\ }\emph{product}\textbf{\emph{\ }}of%
\textbf{\ }\emph{separate}\textbf{\ }sub-system probabilities - which could
be either as $P(\alpha _{1},\alpha _{2}|\Omega _{A1},\Omega _{A2};\lambda
)P(\beta _{1},\beta _{2}|\Omega _{B1},\Omega _{B2};\lambda )$ or as $%
P(\alpha _{j}|\Omega _{A1},\lambda )P(\alpha _{k}|\Omega _{A2},\lambda
)P(\beta _{l}|\Omega _{B1},\lambda )P(\beta _{m}|\Omega _{B2},\lambda )$.
This issue is discussed in an earlier paper on the CGLMP inequalities \cite%
{Dalton19a}. It was pointed out there that a theorem by Fine \cite{Fine82a}
shows that the\textbf{\ }\emph{marginal probabilities}\textbf{\ (}such as $%
P(\alpha _{1},\beta _{1}|\Omega _{A1},\Omega _{B1})$\ in Eq (\ref%
{Eq.MargProb2}) can be written in LHVT\ form. As these marginal
probabilities are used to evaluate the quantity $S$ (see Eq (\ref%
{Eq.CGLMPquantityS}) ) in the CGLMP\ Bell inequality, it\textbf{\ }\emph{%
could}\textbf{\ }be claimed that CGLMP\ is based on \ LHVT and hence a
violation of the inequality shows quantum theory can not be underpinned by
LHVT. That may be the case, but what we\textbf{\emph{\ }}\emph{now}\textbf{\ 
}show is that the CGLMP Bell inequality can\textbf{\ }\emph{also}\textbf{\ b}%
e established using NLHVT (just as for CHSH) - and hence the inequality -
being inconsistent with quantum theory - shows that quantum theory is\textbf{%
\ }\emph{also}\textbf{\ }not underpinned by this non-local hidden variable
theory.

The joint probability $P(\alpha _{j},\alpha _{k},\beta _{l},\beta
_{m}|\Omega _{A1},\Omega _{A2},\Omega _{B1},\Omega _{B2})$ in the CGLMP\
Bell inequality (\ref{Eq.CGLMPquantityS}) is applied in determining the
overall result for four different sets of measurements, each involving one
observable for each sub-system. These measurement choices are ($\Omega
_{A1},\Omega _{B1}$), ($\Omega _{A2},\Omega _{B1}$), ($\Omega _{A2},\Omega
_{B2}$) and ($\Omega _{A1},\Omega _{B2}$).

For completeness - using the\textbf{\ }\emph{locality }condition (\ref%
{Eq.NLHVTCGLMP})for $P(\alpha _{j},\alpha _{k},\beta _{l},\beta _{m}|\Omega
_{A1},\Omega _{A2},\Omega _{B1},\Omega _{B2};\lambda )$\ , we see that 
\textbf{t}he LHVT expression for the joint probability is given by 
\begin{eqnarray}
&&P(\alpha _{j},\alpha _{k},\beta _{l},\beta _{m}|\Omega _{A1},\Omega
_{A2},\Omega _{B1},\Omega _{B2})  \nonumber \\
&=&\dsum\limits_{\lambda }P(\lambda |c)P(\alpha _{j}|\Omega _{A1},\lambda
)P(\alpha _{k}|\Omega _{A2},\lambda )P(\beta _{l}|\Omega _{B1},\lambda
)P(\beta _{m}|\Omega _{B2},\lambda )  \label{Eq.LocalHVTCGLMP}
\end{eqnarray}%
The \emph{marginal probabilities} for \emph{joint} outcomes of \emph{one}
observable for \emph{each} sub-system satisfy \emph{LHVT} conditions \cite%
{Fine82a}. A typical marginal probability for LHVT\ is $P(\alpha _{j},\beta
_{l}|\Omega _{A1},\Omega _{B1})$ $=\dsum\limits_{\lambda }P(\lambda
|c)P(\alpha _{j}|\Omega _{A1},\lambda )P(\beta _{l}|\Omega _{B1},\lambda )$
- which is obviously of LHVT\ form. Note that $\dsum\limits_{\alpha
_{k}\beta _{m}}P(\alpha _{k}|\Omega _{A2},\lambda )P(\beta _{m}|\Omega
_{B2},\lambda )=1$. For the NLHVT\ the results for the marginal probabilites
are just given by (\ref{Eq.MargProb2}).

Bell inequalities involve the \emph{marginal probabilities} that \emph{two}
observables have \emph{same} outcome, such as $P(\Omega _{A1}{\small =}%
\Omega _{B1})=\sum_{j=0}^{d-1}P(\alpha _{j},\beta _{j}|\Omega _{A1},\Omega
_{B1})$, or two observables have outcomes \emph{shifted} (mod$d$), such as $%
P(\Omega _{B1}{\small =}\overline{\Omega _{A2}+1})=\sum_{k=0}^{d-1}P(\alpha
_{k},\beta _{k+1(\func{mod}d)}|\Omega _{A2},\Omega _{B1})$.. The shift by $%
\func{mod}d$ is so that $k+1(\func{mod}d)$ still lies in the range $%
0,1,...,d-1$.

A \emph{typical} CGLMP\ Bell inequality is

\begin{equation}
S=P(\Omega _{A1}{\small =}\Omega _{B1})+P(\Omega _{B1}{\small =}\overline{%
\Omega _{A2}{\small +}1})+P(\Omega _{A2}{\small =}\Omega _{B2})+P(\Omega
_{B2}{\small =}\Omega _{A1})\leq 3  \label{Eq.CHSHBelllneq1}
\end{equation}%
\medskip

\subsection{NLHVT\ Proof of CHSH Bell Inequality}

\label{SubsectProof of CHSH Bell Inequality}

The proof of the CHSH Bell inequality based on local hidden variable theory
is well known \cite{Clauser69a}, and is set out in Ref \ \cite{Durr20a}.
However, as the CHSH Bell inequality involves \emph{two} observables per
sub-system $\Omega _{A1},\Omega _{A2}$ \ for sub-system $A$ with outcomes $%
\alpha _{1},\alpha _{2}$ - and similarly for sub-system $B,$ it is of
interest to consider the situation where a general non-local hidden variable
theory applies. Here there is a \textit{non-local hidden variable theory}
joint probability $P(\alpha _{1},\alpha _{2},\beta _{1},\beta _{2}|\Omega
_{A1},\Omega _{A2},\Omega _{B1},\Omega _{B2})$ for the measurement of \emph{%
all four} observables - even though each pair of sub-system observable
cannot be simultaneously measured according to quantum theory. Probability
conservation gives $\sum_{\alpha _{1},\alpha _{2},\beta _{1},\beta
_{2}}P(\alpha _{1},\alpha _{2},\beta _{1},\beta _{2}|\Omega _{A1},\Omega
_{A2},\Omega _{B1},\Omega _{B2})$ $=1$. The question is: Does the CHSH\ Bell
inequality still apply ?.

In this case for a probabilistic HVT

\begin{eqnarray}
&&P(\alpha _{1},\alpha _{2},\beta _{1},\beta _{2}|\Omega _{A1},\Omega
_{A2},\Omega _{B1},\Omega _{B2}) \\
&=&\dsum\limits_{\lambda }P(\lambda |c)P(\alpha _{1},\alpha _{2},\beta
_{1},\beta _{2}|\Omega _{A1},\Omega _{A2},\Omega _{B1},\Omega _{B2};\lambda )
\end{eqnarray}%
The \textit{marginal probabilities }for measurement outcomes for \textit{one}
observable for \emph{each} sub-system would be 
\begin{eqnarray}
P(\alpha _{1},\beta _{1}|\Omega _{A1},\Omega _{B1}) &=&\dsum\limits_{\lambda
}P(\lambda |c)\dsum\limits_{\alpha _{2}\beta _{2}}P(\alpha _{1},\alpha
_{2},\beta _{1},\beta _{2}|\Omega _{A1},\Omega _{A2},\Omega _{B1},\Omega
_{B2};\lambda )  \nonumber \\
P(\alpha _{2},\beta _{1}|\Omega _{A2},\Omega _{B1}) &=&\dsum\limits_{\lambda
}P(\lambda |c)\dsum\limits_{\alpha _{1}\beta _{2}}P(\alpha _{1},\alpha
_{2},\beta _{1},\beta _{2}|\Omega _{A1},\Omega _{A2},\Omega _{B1},\Omega
_{B2};\lambda )  \nonumber \\
P(\alpha _{1},\beta _{2}|\Omega _{A2},\Omega _{B2}) &=&\dsum\limits_{\lambda
}P(\lambda |c)\dsum\limits_{\alpha _{2}\beta _{1}}P(\alpha _{1},\alpha
_{2},\beta _{1},\beta _{2}|\Omega _{A1},\Omega _{A2},\Omega _{B1},\Omega
_{B2};\lambda )  \nonumber \\
P(\alpha _{2},\beta _{2}|\Omega _{A1},\Omega _{B2}) &=&\dsum\limits_{\lambda
}P(\lambda |c)\dsum\limits_{\alpha _{1}\beta _{1}}P(\alpha _{1},\alpha
_{2},\beta _{1},\beta _{2}|\Omega _{A1},\Omega _{A2},\Omega _{B1},\Omega
_{B2};\lambda )  \nonumber \\
&&
\end{eqnarray}%
where all \emph{four} choices of the pairs of\textbf{\ }sub-system
observables $(\Omega _{Aa},\Omega _{Bb})$ are considered. Note the sums over
the \emph{unrecorded} observable outcomes. These expressions are as (\ref%
{Eq.MargProb2}) with a notation change.

Typical terms in the expression for $S$ can then be obtained. For the first
term 
\begin{eqnarray}
\left\langle \Omega _{A1}\otimes \Omega _{B1}\right\rangle _{LHVT}
&=&\sum_{\alpha _{1},\beta _{1},}\{\alpha _{1}\beta _{1}\}P(\alpha
_{1},\beta _{1}|\Omega _{A1},\Omega _{B1})  \nonumber \\
&=&\dsum\limits_{\lambda }P(\lambda |c)\sum_{\alpha _{1},\alpha _{2},\beta
_{1},\beta _{2}}P(\alpha _{1},\alpha _{2},\beta _{1},\beta _{2}|\Omega
_{A1},\Omega _{A2},\Omega _{B1},\Omega _{B2};\lambda )\{\alpha _{1}\beta
_{1}\}  \nonumber \\
&&
\end{eqnarray}%
The other terms are derived similarly.

The quantity $S$ in the CHSH inequality would then be given by 
\begin{eqnarray}
S &=&\dsum\limits_{\lambda }P(\lambda |c)\sum_{\alpha _{1},\alpha _{2},\beta
_{1},\beta _{2}}P(\alpha _{1},\alpha _{2},\beta _{1},\beta _{2}|\Omega
_{A1},\Omega _{A2},\Omega _{B1},\Omega _{B2};\lambda )\{\alpha _{1}\beta
_{1}+\alpha _{1}\beta _{2}+\alpha _{2}\beta _{1}-\alpha _{2}\beta _{2}\} 
\nonumber \\
&=&\dsum\limits_{\lambda }P(\lambda |c)\sum_{\alpha _{1},\alpha _{2},\beta
_{1},\beta _{2}}P(\alpha _{1},\alpha _{2},\beta _{1},\beta _{2}|\Omega
_{A1},\Omega _{A2},\Omega _{B1},\Omega _{B2};\lambda )\{\alpha _{1}(\beta
_{1}+\beta _{2})+\alpha _{2}(\beta _{1}-\beta _{2})\}  \nonumber \\
&&  \label{Eq.NLHVTResultSinCHSH}
\end{eqnarray}

In the CHSH case where $\Omega _{A1},\Omega _{A2},\Omega _{B1},\Omega _{B2}$
are all components of \textit{Pauli spin observables} in spin $1/2$
sub-systems, all the \emph{outcomes} are either $+1$ or $-1$. In which case $%
(\beta _{1}+\beta _{2})$ can either be $-2$, $0$, $0$ or $+2$ and for these
situations $(\beta _{1}-\beta _{2})$ will be $0$, $+2$, $-2$ or $0$
respectively. Thus \emph{one} of the two factors $(\beta _{1}+\beta _{2})$
or $(\beta _{1}-\beta _{2})$ will be zero whilst the other has a magnitude
of $2.$ The magnitude of both $\alpha _{1}$ and $\alpha _{2}$ is $1.$ Thus
in all cases $|\{\alpha _{1}(\beta _{1}+\beta _{2})+\alpha _{2}(\beta
_{1}-\beta _{2})\}|\;=2$.

Hence 
\begin{eqnarray}
|S|\, &\leq &\,\dsum\limits_{\lambda }P(\lambda |c)\sum_{\alpha _{1},\alpha
_{2},\beta _{1},\beta _{2}}P(\alpha _{1},\alpha _{2},\beta _{1},\beta
_{2}|\Omega _{A1},\Omega _{A2},\Omega _{B1},\Omega _{B2};\lambda )\times 2 
\nonumber \\
&\leq &\,2\text{. }
\end{eqnarray}%
as required - using the magnitude of a sum being less than or equal to the
sum of the magnitudes.

The CHSH\ Bell inequality can also be proved based on the LHVT form (\ref%
{Eq.LocalHVTCGLMP})\ for the joint probability with\textbf{\ }\emph{two }%
observables per sub-system, where $P(\alpha _{1},\alpha _{2},\beta
_{1},\beta _{2}|\Omega _{A1},\Omega _{A2},\Omega _{B1},\Omega _{B2};\lambda
)=P(\alpha _{1}|\Omega _{A1},\lambda )P(\alpha _{2}|\Omega _{A2},\lambda
)P(\beta _{1}|\Omega _{B1},\lambda )P(\beta _{2}|\Omega _{B2},\lambda ).$ So
it is only for the\textbf{\ }\emph{one}\textbf{\ }observable per sub-system
form (\ref{Eq.NLHVTJointProb}) for the NLHVT where the CHSH\ Bell inequality
proof fails.

Thus the CHSH inequality\textit{\ also} holds for this type of NLHVT, so its
violation for the singlet state listed in Section \ref{SubSysCHSHTest} shows
that this type of \textit{non-localit}y is \textit{not consistent} with
quantum theory. Hence we find the new result that quantum theory \emph{cannot%
} be underpinned by this type of NLHVT. Evidently, the violation of the CHSH
Bell inequality also rules out a very general type of NLHVT as well ! We
will find that the same outcome applies when the CGLMP\ Bell inequalities
are considered. \smallskip

\subsection{Observables and Quantum State}

We now introduce the quantum theory observables that will be used in the
CGLMP inequalities, along with the quantum state which will be used for
comparison with HVT predictions - both local and non-local. .

The Bell inequality (\ref{Eq.CHSHBelllneq1}) based on the LHVT expression is 
\emph{violated} for the quantum theory\textbf{\ }\emph{maximally entangled} 
\emph{state} 
\begin{equation}
\left\vert \Psi \right\rangle =\frac{1}{\sqrt{d}}\sum_{j=0}^{d-1}\left\vert
j\right\rangle _{A}\otimes \left\vert j\right\rangle _{B}
\label{Eq.MaxEntState}
\end{equation}%
for quantum theory\textbf{\ }\emph{observables} whose \emph{eigenstates} for 
\emph{eigenvalues} $k$, $l$ in the case of observables $\Omega _{Aa}$ , $%
\Omega _{Bb}$ ($a,b=1,2$) are: 
\begin{eqnarray}
\left\vert k\right\rangle _{A,a}\, &{\small =}&\frac{1}{\sqrt{d}}%
\sum_{j=0}^{d-1}\exp i\frac{2\pi }{d}j\left( k+\theta _{a}\right)
\;\left\vert j\right\rangle _{A}\qquad a={\small 1,2}\quad {\small \theta }%
_{1}={\small 0},{\small \theta }_{2}={\small 1/2},
\label{Eq.EigenvectorsObserA} \\
\left\vert l\right\rangle _{B,b}\, &{\small =}&\frac{1}{\sqrt{d}}%
\sum_{j=0}^{d-1}\exp i\frac{2\pi }{d}j\left( -l+\phi _{b}\right)
\;\left\vert j\right\rangle _{B}\qquad b={\small 1,2}\quad {\small \phi }%
_{1}={\small 1/4},{\small \phi }_{2}={\small -1/4}.  \nonumber \\
&&  \label{Eq.EigenvectorsObservB}
\end{eqnarray}%
\smallskip Here the $\left\vert j\right\rangle _{A}$ and $\left\vert
j\right\rangle _{B}$ are orthonormal basis states for sub-systems $A$, $B$
respectively. Note that the observables $\Omega _{A1}$, $\Omega _{A2}$ are 
\emph{not} assumed to \emph{commute}, neither are $\Omega _{B1}$, $\Omega
_{B2}$. These observables are alo \emph{not} required to \ have any \emph{%
obvious} physical meaning. However the eigenvectors $\left\vert
k\right\rangle _{A,a}$, $\left\vert l\right\rangle _{B,b}$ can be seen to
satisfy the expected \emph{orthonormality conditions} for different $k$, $l$%
. Thus 
\begin{eqnarray}
\left\langle k\;|k^{\ast }\right\rangle _{A,a} &=&\frac{1}{d}%
\sum_{j=0}^{d-1}\exp i\frac{2\pi }{d}j\left( -k+k^{\ast }\right)  \nonumber
\\
&=&\frac{1}{d}\frac{(1-\exp i\frac{2\pi }{d}d(-k+k^{\ast }))}{(1-\exp i\frac{%
2\pi }{d}(-k+k^{\ast }))}=0\qquad if\text{ }k\neq k^{\ast }  \nonumber \\
&=&1\qquad if\text{ }k=k^{\ast }
\end{eqnarray}%
with a similar result for $\left\langle l\;|l^{\ast }\right\rangle _{B,b%
\text{ .}}$Bell non-locality for \emph{above} CGLMP Bell inequalitity (\ref%
{Eq.CHSHBelllneq1})\ \emph{occurs} for $d=2$.

The basic \emph{quantum} expressions for the \emph{sub-system probabilities}
in the case of the \emph{maximally entangled state} is (see \cite{Collins02a}%
) 
\begin{eqnarray}
P(\alpha _{k},\beta _{l}|\Omega _{Aa},\Omega _{Bb},\rho )_{Q}
&=&|(\left\langle k\;|_{A,a}\otimes \right\langle l\;|_{B,b}|)\;\left\vert
\Psi \right\rangle |^{2}  \nonumber \\
&=&\frac{1}{d}\;|\,\sum_{j=0}^{d-1}\left\langle k\;|_{A,a}|\;j\right\rangle
_{A}\;|\left\langle l\;|_{B,b}|\;j\right\rangle _{B}|^{2}  \nonumber \\
&=&\frac{1}{d^{3}}|\;\sum_{j=0}^{d-1}\exp -i\frac{2\pi }{d}j(k+\theta
_{a})\;\exp -\frac{2\pi }{d}j(-l+\phi _{b})|^{2}  \nonumber \\
&=&\frac{1}{d^{3}}\;\frac{|(1-\exp (-i2\pi (k-l+\theta _{A}+\phi _{B}))|^{2}%
}{|(1-\exp (-i\frac{2\pi }{d}(k-l+\theta _{A}+\phi _{B}))|^{2}}  \nonumber \\
&=&\frac{1}{d^{3}}\;\frac{\sin ^{2}(\pi (k-l+\theta _{a}+\phi _{b}))}{\sin
^{2}(\frac{\pi }{d}(k-l+\theta _{a}+\phi _{b}))}  \nonumber \\
&=&\frac{1}{2d^{3}}\;\frac{1}{\sin ^{2}(\frac{\pi }{d}(k-l+\theta _{a}+\phi
_{b}))}  \label{Eq.QThyResultJointProb}
\end{eqnarray}%
where we have used $\sin ^{2}(\pi (k-l+\theta _{a}+\phi _{b}))=\{\sin \pi
(k-l)\cos \pi (\theta _{a}+\phi _{b})+(\cos \pi (k-l)\sin \pi (\theta
_{a}+\phi _{b})\}^{2}=\{0+(\pm 1)(\pm 1/\sqrt{2})\}^{2}$, noting that the
only $\theta _{a}+\phi _{b}$ values are $1/4,-1/4,3/4,1/4$ and $\sin \pi
(1/4)=-\sin \pi (-1/4)=\sin \pi (3/4)=1/\sqrt{2}$. \textbf{\ }Also, $(k-l)$
is always an integer.

Macroscopic Bell non-locality \emph{occurs} in \emph{other} CGLMP Bell
inequalities for $d$ large. Experimental verification up to $d=12$ was found
by \cite{Dada11a}{\small . }\bigskip \pagebreak

\subsection{NLHVT Case}

\label{SubSectionNLHVTCaseCGLMP}

We now consider a $\emph{more}$ $\emph{general}$\emph{\ CGLMP Bell inequality%
} involving \ 
\begin{equation}
S=P(\Omega _{A1}{\small =}\overline{\Omega _{B1}+\Delta _{11}})+P(\Omega
_{B1}{\small =}\overline{\Omega _{A2}{\small +}\Delta _{12}})+P(\Omega _{A2}%
{\small =}\overline{\Omega _{B2}+\Delta _{22}})+P(\Omega _{B2}{\small =}%
\overline{\Omega _{A1}+\Delta _{21})}  \label{Eq.CGLMPquantityS}
\end{equation}%
where the outcome shifts\textbf{\ }$\Delta _{ij}$ are all integers and it is
understood that $+\Delta _{ij}$ means $+\Delta _{ij}(\func{mod}d)$.

As in the LHVT case, there are two \emph{observables} $\Omega _{A1},\Omega
_{A2}$ for sub-system $A$ and two observables $\Omega _{B1},\Omega _{B2}$
for sub-system $B$. The \emph{eigenvectors} for all the observables are (\ref%
{Eq.EigenvectorsObserA}), (\ref{Eq.EigenvectorsObservB})\textbf{\ }\ as for
the LHVT case. We will also consider quantum probabilities for the same 
\emph{maximally entangled state}\textbf{\ }(\ref{Eq.MaxEntState})\ that was
treated in the LHVT case. However, \ we now use the NLHVT expressions for
the probabilities and no longer invoke\textbf{\ }\emph{locality.}\textbf{\ }

First we construct the NLHVT expression for $S$. We have using the%
{\footnotesize \ }\emph{marginal probabilities}\textbf{\ }in (\ref%
{Eq.MargProb2})%
\begin{eqnarray}
S &=&\dsum\limits_{j,l}P(\alpha _{j},\beta _{l}|\Omega _{A1},\Omega
_{B1})\,\delta _{j,l+\Delta _{11}}+\dsum\limits_{k,l}P(\alpha _{k},\beta
_{l}|\Omega _{A2},\Omega _{B1})\,\delta _{l,k+\Delta _{12}}  \nonumber \\
&&+\dsum\limits_{k,m}P(\alpha _{k},\beta _{m}|\Omega _{A2},\Omega
_{B2})\,\delta _{k,m+\Delta _{22}}+\dsum\limits_{j,m}P(\alpha _{j},\beta
_{m}|\Omega _{A1},\Omega _{B2})\,\delta _{m,j+\Delta _{21}}  \nonumber \\
&=&\dsum\limits_{\lambda }P(\lambda |c)\left\{ 
\begin{array}{c}
\dsum\limits_{j,l}\dsum\limits_{\alpha _{k}\beta _{m}}P(\alpha _{j},\alpha
_{k},\beta _{l},\beta _{m}|\Omega _{A1},\Omega _{A2},\Omega _{B1},\Omega
_{B2};\lambda )\;\,\delta _{j,l+\Delta _{11}} \\ 
+\dsum\limits_{k,l}\dsum\limits_{\alpha _{j}\beta _{m}}P(\alpha _{j},\alpha
_{k},\beta _{l},\beta _{m}|\Omega _{A1},\Omega _{A2},\Omega _{B1},\Omega
_{B2};\lambda )\;\,\delta _{l,k+\Delta _{12}} \\ 
+\dsum\limits_{k,m}\dsum\limits_{\alpha _{j}\beta _{l}}P(\alpha _{j},\alpha
_{k},\beta _{l},\beta _{m}|\Omega _{A1},\Omega _{A2},\Omega _{B1},\Omega
_{B2};\lambda )\;\,\delta _{k,m+\Delta _{22}} \\ 
+\dsum\limits_{j,m}\dsum\limits_{\alpha _{k}\beta _{l}}P(\alpha _{j},\alpha
_{k},\beta _{l},\beta _{m}|\Omega _{A1},\Omega _{A2},\Omega _{B1},\Omega
_{B2};\lambda )\,\delta _{m,,j+\Delta _{21}}%
\end{array}%
\right\}  \nonumber \\
&=&\dsum\limits_{\lambda }P(\lambda |c)\left\{ 
\begin{array}{c}
\dsum\limits_{jklm}P(\alpha _{j},\alpha _{k},\beta _{l},\beta _{m}|\Omega
_{A1},\Omega _{A2},\Omega _{B1},\Omega _{B2};\lambda )\; \\ 
\times \{\delta _{j,l+\Delta _{11}}+\delta _{l,k+\Delta _{12}}+\delta
_{k,m+\Delta _{22}}+\delta _{m,j+\Delta _{21}}\}%
\end{array}%
\right\}  \label{Eq.SforNLHVTNew}
\end{eqnarray}

Clearly the quantity $\delta =\{\delta _{j,l+\Delta _{11}}+\delta
_{l,k+\Delta _{12}}+\delta _{k,m+\Delta _{22}}+\delta _{m,j+\Delta _{21}}\}$
can have values $4,3,2,1$ depending on how many of the Kronecka deltas are
equal to $1$ rather than \ $0$. We\textbf{\ }see that the CGLMP Bell
inequality in \ the NLHVT case will just depend on the value for $\delta $\ 
\begin{eqnarray}
S &\leq &\dsum\limits_{\lambda }P(\lambda |c)\left\{
\dsum\limits_{jklm}P(\alpha _{j},\alpha _{k},\beta _{l},\beta _{m}|\Omega
_{A1},\Omega _{A2},\Omega _{B1},\Omega _{B2};\lambda )\;\times \delta
\right\}  \nonumber \\
&\leq &\delta  \label{Eq.IneqalCGLMP}
\end{eqnarray}%
using $\dsum\limits_{j,k,l,m}P(\alpha _{j},\alpha _{k},\beta _{l},\beta
_{m}|\Omega _{A1},\Omega _{A2},\Omega _{B1},\Omega _{B2};\lambda )=1$ and $%
\dsum\limits_{\lambda }P(\lambda |c)=1$. In the next Section we determine%
\textbf{\ }\emph{sufficiency requirements}\textbf{\ }on the\textbf{\ }\emph{%
shifts }$\Delta _{ij}$ for the cases\textbf{\ }$\delta =4,3,2,1$.

By comparison with (\ref{Eq.SforNLHVTNew})\ the\textbf{\ }correspondiing
expression \textbf{if} LHVT\ applies\textbf{\ }is 
\begin{eqnarray}
S &=&\dsum\limits_{\lambda }P(\lambda |c)\left\{ 
\begin{array}{c}
\dsum\limits_{j,k,l,m}P(\alpha _{j}|\Omega _{A1},\lambda )P(\alpha
_{k}|\Omega _{A2},\lambda )P(\beta _{l}|\Omega _{B1},\lambda )P(\beta
_{m}|\Omega _{B2},\lambda ) \\ 
\times (\delta _{j,l+\Delta _{11}}+\delta _{l,k+\Delta _{12}}+\delta
_{k,m+\Delta _{22}}+\delta _{m,j+\Delta _{21}})%
\end{array}%
\right\}  \nonumber \\
&&  \label{Eq.SforLHVT}
\end{eqnarray}%
where here the locality based\textbf{\ }\emph{factorisation}\textbf{\ }of $%
P(\alpha _{j},\alpha _{k},\beta _{l},\beta _{m}|\Omega _{A1},\Omega
_{A2},\Omega _{B1},\Omega _{B2};\lambda )$ occurs. However the\textbf{\ }%
\emph{same}\textbf{\ }sum of Kronecka deltas is present. In this LHVT\
version we see that we also have%
\begin{equation}
S\leq \delta  \label{Eq.InequalCGLMPLocalHVT}
\end{equation}%
using $\dsum\limits_{j,k,l,m}P(\alpha _{j}|\Omega _{A1},\lambda )P(\alpha
_{k}|\Omega _{A2},\lambda )P(\beta _{l}|\Omega _{B1},\lambda )P(\beta
_{m}|\Omega _{B2},\lambda )=1$\ as each factor is equal to $1$, for example%
\textbf{\ }$\sum_{j}P(\alpha _{j}|\Omega _{A1},\lambda )=1$\textbf{. }

This means that if the NLHVT is violated by quantum theory, then the LHVT\
will \emph{also} be violated\textbf{. }

Note that the sum of the joint measurement probabilities where \emph{%
specific relationships} occur between the\emph{\ measurement outcomes} for
the two sub-systems is in general less that the sum of the joint measurement
probabilities for \emph{all} such relationships. Thus $P(\Omega _{A1}{\small %
=}\overline{\Omega _{B1}+\Delta _{11}})=\dsum\limits_{\lambda }P(\lambda
|c)\dsum\limits_{j,l}P(\alpha _{j},\beta _{l}|\Omega _{A1},\Omega
_{B1},\lambda )\,\delta _{j,l+\Delta _{11}}\leq \dsum\limits_{\lambda
}P(\lambda |c)\dsum\limits_{j,l}P(\alpha _{j},\beta _{l}|\Omega _{A1},\Omega
_{B1},\lambda )=P(\Omega _{A1},\Omega _{B1})=1$., where $P(\Omega
_{A1},\Omega _{B1})$ is the overall probability for \emph{all }outcomes of
joint measurements for the choice $\Omega _{A1}$ and $\Omega _{B1}$. Similar
considerations apply to $P(\Omega _{B1}{\small =}\overline{\Omega _{A2}%
{\small +}\Delta _{12}})$, $P(\Omega _{A2}{\small =}\overline{\Omega
_{B2}+\Delta _{22}})$ and $P(\Omega _{B2}{\small =}\overline{\Omega
_{A1}+\Delta _{21})\text{.}}$. \bigskip

\pagebreak

\section{REQUIREMENTS FOR\ NLHVT\ BELL\ INEQUALITIES}

First we consider the \emph{conditions} involving the Kroneka deltas
involved in the simple CGLMP inequality and the \emph{matrix approach} used
to treat them. Since the requirements \textbf{\emph{only} }involve
considering the situations where the Kronecka sum $\delta $ is given by $%
4,3,2$ or $1$, it follows that the requirements will be the same for both
the LHVT and NLHVT cases.\textbf{\ \medskip }

\subsection{Conditions and Basic Matrix Approach}

To derive the \emph{sufficiency requirements}\textbf{\ }on the\textbf{\ }%
\emph{outcome shifts }$\Delta _{ij}$ for the NLHVT Bell Inequalities (\ref%
{Eq.IneqalCGLMP}) we must consider the \emph{conditions} where the various
Kronecka deltas are equal to \emph{unity}. These conditions are

\begin{eqnarray}
C1 &:&\qquad j=l+\Delta _{11}  \nonumber \\
C2 &:&\qquad l=k+\Delta _{12}  \nonumber \\
C3 &:&\qquad k=m+\Delta _{22}  \nonumber \\
C4 &:&\qquad m=j+\Delta _{21}
\end{eqnarray}%
We will consider \emph{cases} where; Case 1 -\emph{\ all} of $C1,C2,C3,C4$
are \emph{true (T)}; Case 2 - \emph{three} of $C1,C2,C3,C4$ are true and 
\emph{one} is\emph{\ false (F)}; Case 3 - \emph{two} of $C1,C2,C3,C4$ are
true and \emph{two} are false; Case 4 - \emph{one} of $C1,C2,C3,C4$ is true
and \emph{three} are false. For these four Cases, there are a number of%
\textbf{\ }\emph{Sub-Cases}\textbf{\ }depending on which of\textbf{\ }$%
C1,C2,C3,C4$ are \emph{true }and which are\textbf{\ }\emph{false}\textbf{\ - 
}one for Case 1, four for Case 2, six for Case 3 and four for Case 4. For
the four Cases the value of $\delta =$\ $(\delta _{j,l+\Delta _{11}}+\delta
_{l,k+\Delta _{12}}+\delta _{k,m+\Delta _{22}}+\delta _{m,j+\Delta _{21}})$
are the same for every Sub-Case, namely $\delta =4,3,2,1$.\ for Cases $%
1,2,3,4$\ respectively. As described in Sect.\ref{SubSectionNLHVTCaseCGLMP}
for these four Cases the values of $\delta $ are used to determine the Bell
inequality $S\leq \delta $\textbf{\ }(\ref{Eq.IneqalCGLMP})\ using Eq. (\ref%
{Eq.SforNLHVTNew}) for $S$ (defined in Eq. (\ref{Eq.CGLMPquantityS})).
However, the\emph{\ requirements} on the\textbf{\ }\emph{shifts}\textbf{\ }$%
\Delta _{ij}$ for each\textbf{\ }\emph{Sub-Case}\textbf{\ }will be\textbf{\ }%
\emph{different}\textbf{\emph{,} }and these are set out below. Details are
given in Appendix B. .

To derive these requirements it turns out that we must consider matrix
equations of the form \ 
\begin{equation}
M\;\mathbf{v=\Delta }  \label{Eq.MatrixEqns}
\end{equation}%
where $M$ is a square matrix and $\mathbf{v}$ is a column vector whose
elements are some (or all) of the integers $j,k,l,m$ and $\mathbf{\Delta }$
is a column vector whose elements are sums of the various $\Delta _{ij}$. We
will use the result that if the \emph{determinent} of $M$ is zero ($|M|=0$)
then there is only a possible solution for $\mathbf{v}$ if \emph{all} the
elements of the corresponding $\mathbf{\Delta }$ are zero. This will then
lead to a set of \emph{requirements} on the shifts $\Delta _{ij\text{ }}$to
lead to Bell inequalities: $S\leq 4$, $S\leq 3$, $S\leq 2$, $S\leq 1$ for
each Sub-Case of the Cases 1, 2, 3, 4. Note that we merely have\ to have%
\textbf{\ \emph{one} }element of $\Delta $ \textbf{to be }\emph{non-zero}%
\textbf{\ }to show there is\textbf{\ \emph{no} }solution for $\mathbf{v}$.

The key point is that the CGLMP Bell Inequalities are of the form $S\leq
\delta $, so they\textbf{\ }\emph{only}\textbf{\ }depend on the\textbf{\ }%
\emph{number}\textbf{\ }of the conditions $C1,C2,C3,C4$ that are satisfied.
The\textbf{\ }\emph{basic strategy}\textbf{\ }for each Sub Case is to take
each condition that is required to be\textbf{\ }\emph{False}\textbf{, t}hen
substitute the conditions that are required to be \emph{True} into these to
arrive at the\textbf{\ }\emph{matrix equation}\textbf{\ }(\ref{Eq.MatrixEqns}%
) for the\textbf{\ }\emph{remaining}\textbf{\ }indices from $j,k,l,m$\ and
then\textbf{\ }\emph{identify}\textbf{\ }the column vector $\Delta $ for
this \ Sub-Case\textbf{.} We then show that the determinent $|M|$\ is zero.
As then there is to be\textbf{\ }\emph{no solution} for the conditions that
are\textbf{\ }\emph{False}\textbf{, }we then obtain the requirement that%
\textbf{\ }\emph{at least one }of \ the elements of $\Delta $\ is non-zero.
Hence the requirement for each Sub-Case will be a list of\textbf{\ }\emph{%
And/Or}\textbf{\ }statements. Each statement is a set of\textbf{\ }$\emph{%
And/Or}$ \emph{inequalities}\textbf{\ }(or an equality in the Case of\textbf{%
\ }$S\leq 4$) arranged in\textbf{\ }\emph{rows}\textbf{\ }for the various
Sub-Cases, and involving the\textbf{\ }\emph{outcome shifts}\textbf{\ }$%
(\Delta _{11},\Delta _{12},\Delta _{22},\Delta _{21})$\textbf{. }If\textbf{\ 
\emph{at }}\emph{least one}\textbf{\ }of these these inequalities are
satisfied in\emph{\ every row}\textbf{, }then we have obtained\textbf{\ }%
\emph{sufficiency requirements} to show \medskip $S\leq \delta $\ for the
Case involved. Note that we do not obtain\textbf{\ }\emph{necessity
requirements, }as a particular set of outcome shifts $(\Delta _{11},\Delta
_{12},\Delta _{22},\Delta _{21})$ may provide sufficiency requirements for%
\textbf{\ }\emph{more than one}\textbf{\ }of\textbf{\ }$S\leq 3$, $S\leq 2$%
\textbf{\ }and\textbf{\ } $S\leq 1$. \ If more than one is satisfied then the%
\textbf{\ }\emph{overall}\textbf{\ }NLHVT prediction is $S$ $\leq $ the%
\textbf{\ }\emph{smallest}\textbf{\ }of\textbf{\ }$3,2,1$\ satisfied. This
is logical since if for the outcome shifts $(\Delta _{11},\Delta
_{12},\Delta _{22},\Delta _{21})$ we can show that $S\leq 3$, $S\leq 2$\ and 
$S\leq 1$ then overall we have $S\leq 1$. \ The Case of $S\leq 4$ is an \
exception, as we will see. Sufficiency requirements for $S\leq 4$ are
incompatible with those for\textbf{\ }$S\leq 3$, $S\leq 2$\textbf{\ }and%
\textbf{\ } $S\leq 1$. \medskip

\subsection{Requirements for Bell Inequality Tests}

These Sub-Case requirements are derived in Appendix B for all Sub-Cases of
the four Cases. These are summarised below in Eqs. (\ref{Eq.RequireSLessFour}%
),\ (\ref{Eq.RequireSLessThree}) , (\ref{Eq.RequireLessTwoSubCases}) and (%
\ref{Eq.RequireLessOneSubCases}). The \emph{resulting requirements} for the
four Cases is based on the distinct Sub-Case requirements and\emph{\ must}%
\textbf{\ }cover all Sub-Cases to\textbf{\ }constitute \emph{sufficient}
(rather than \emph{necessary}) requirements for concluding that according to
our general NLHVT we have $S$ satisfying $S\leq 4$, $S\leq 3$, $S\leq 2$ or $%
S\leq 1$. These requirements are set out below in Eqs. (\ref%
{Eq.RequireSLessFour}), (\ref{Eq.RequireSLessThree}), (\ref%
{Eq.RequireLessTwoSubCases}) and (\ref{Eq.RequireLessOneSubCases})
respectively. In deriving these results we have used $\dsum%
\limits_{j,k,l,m}P(\alpha _{j},\alpha _{k},\beta _{l},\beta _{m}|\Omega
_{A1},\Omega _{A2},\Omega _{B1},\Omega _{B2};\lambda )=1$ and $%
\dsum\limits_{\lambda }P(\lambda |c)=1$.\medskip

\subsubsection{Case1 - S$\leq 4$}

The requirements for the single Sub-Case\textbf{\ }are

\begin{equation}
\Delta _{11}=0\quad And\quad \Delta _{12}=0\quad And\quad \Delta
_{22}=0\quad And\quad \Delta _{21}=0  \label{Eq.RequireSLessFour}
\end{equation}

If \emph{all} these requirements are satisfied the Bell inequality $S\leq 4$
would be guaranteed. Note that if these requirements are satisfied then
those for $S\leq 3,S\leq 2,S\leq 1$ are not.\medskip

\subsubsection{Case 2 - S $\leq 3$}

The requirements for all four Sub-Cases are the same 
\begin{equation}
\Delta _{12}+\Delta _{11}+\Delta _{21}+\Delta _{22}\neq 0\qquad Sub-Cases%
\text{ }a,b,c,d  \label{Eq.RequireSLessThree}
\end{equation}

If this requirement applies the Bell inequality $S\leq 3$\ would be
satisfied. \medskip

\subsubsection{Case 3 - S $\leq 2$}

The requirements for the six different Sub-Cases are%
\begin{eqnarray}
\Delta _{22} &\neq &0\quad And/Or\quad \Delta _{12}+\Delta _{11}+\Delta
_{21}\neq 0\qquad Sub-Case\text{ }a  \nonumber \\
\Delta _{22}+\Delta _{12} &\neq &0\quad And/Or\quad \Delta _{11}+\Delta
_{21}\neq 0\qquad \qquad Sub-Case\text{ }b  \nonumber \\
\Delta _{12} &\neq &0\quad And/Or\quad \Delta _{11}+\Delta _{21}+\Delta
_{22}\neq 0\qquad Sub-Case\text{ }c  \nonumber \\
\Delta _{22}+\Delta _{12}+\Delta _{11} &\neq &0\quad And/Or\quad \Delta
_{21}\neq 0\qquad \qquad Sub-Case\text{ }d  \nonumber \\
\Delta _{12}+\Delta _{11} &\neq &0\quad And/Or\quad \Delta _{21}+\Delta
_{22}\neq 0\qquad \qquad Sub-Case\text{ }e  \nonumber \\
\Delta _{11} &\neq &0\quad And/Or\quad \Delta _{21}+\Delta _{22}+\Delta
_{12}\neq 0\qquad Sub-Case\text{ }f  \nonumber \\
&&  \label{Eq.RequireLessTwoSubCases}
\end{eqnarray}

Since we wish to \emph{guarantee} that the Bell inequality $S\leq 2$ would
occur irrespective of the Sub-Case, we can be \emph{certain} of this by
checking for a given set of\textbf{\ }\emph{outcome shifts}\textbf{\ }$%
(\Delta _{11},\Delta _{12},\Delta _{22},\Delta _{21})$ that \emph{at least
one} of the \emph{distinct} inequalities in \emph{each row} of the last
equation is satisfied. \ For exanple, the requirements\textbf{\ }$\Delta
_{11}\neq 0,\Delta _{12}\neq 0,\Delta _{22}\neq 0$\textbf{\ }and $\Delta
_{21}\neq 0$ would satisfy the requirements for Sub-Cases $(a),(c),(d)$\ and 
$(f)$ but not those for Sub-Cases $(b)$ and $(e)$, so with these four
requirements\textbf{\ \emph{alone} }we cannot conclude that $S\leq 2$.
\medskip

\subsubsection{Case 4 - S $\leq 1$}

The requirements for the four different Sub-Cases are%
\begin{eqnarray}
\Delta _{12} &\neq &0\quad And/Or\quad \Delta _{22}\neq 0\quad And/Or\quad
\Delta _{11}+\Delta _{21}\neq 0\qquad Sub-Case\text{ }a  \nonumber \\
\Delta _{11}+\Delta _{12} &\neq &0\quad And/Or\quad \Delta _{22}\neq 0\quad
And/Or\quad \Delta _{21}\neq 0\qquad Sub-Case\text{ }b  \nonumber \\
\Delta _{11} &\neq &0\quad And/Or\quad \Delta _{22}+\Delta _{12}\neq 0\quad
And/Or\quad \Delta _{21}\neq 0\qquad Sub-Case\text{ }c  \nonumber \\
\Delta _{11} &\neq &0\quad And/Or\quad \Delta _{12}\neq 0\quad And/Or\quad
\Delta _{22}+\Delta _{21}\neq 0\qquad \qquad Sub-Case\text{ }d  \nonumber \\
&&  \label{Eq.RequireLessOneSubCases}
\end{eqnarray}

Since we wish to \emph{guarantee} that the Bell inequality $S\leq 1$ would
occur irrespective of the sub-case, we can be \emph{certain} of this by
checking for a given set of outcome shifts\textbf{\ }$(\Delta _{11},\Delta
_{12},\Delta _{22},\Delta _{21})$ that \emph{at least one }of\emph{\ } the 
\emph{distinct} inequalities in \emph{each row} of the last equation are
satisfied. \medskip

\subsection{Example}

As an example of what happens for a particular choice of outcome shifts we
choose the case $(\Delta _{11},\Delta _{12},\Delta _{22},\Delta
_{21})=(1,-2,1,-2).$\ It is obvious that the requirement for $S\leq 4$ is
not satisfied.

We substitute the numerical values into Eqs. (\ref{Eq.RequireSLessThree}), (%
\ref{Eq.RequireLessTwoSubCases}) and (\ref{Eq.RequireLessOneSubCases}) .

For $S\leq 3$ (all Sub-Case) we see that $\Delta _{12}+\Delta _{11}+\Delta
_{21}+\Delta _{22}=(-2)\neq 0$ as required.

For $S\leq 2\ $we have\textbf{\ }%
\begin{eqnarray}
\Delta _{22} &=&(2)\neq 0\quad And/Or\quad \Delta _{12}+\Delta _{11}+\Delta
_{21}=(-3)\neq 0\qquad Sub-Case\text{ }a  \nonumber \\
\Delta _{22}+\Delta _{12} &=&(-1)\neq 0\quad And/Or\quad \Delta _{11}+\Delta
_{21}=(-1)\neq 0\qquad \qquad Sub-Case\text{ }b  \nonumber \\
\Delta _{12} &\neq &(-2)\neq 0\quad And/Or\quad \Delta _{11}+\Delta
_{21}+\Delta _{22}=(0)\neq 0\qquad Sub-Case\text{ }c  \nonumber \\
\Delta _{22}+\Delta _{12}+\Delta _{11} &=&(0)\neq 0\quad And/Or\quad \Delta
_{21}=(-2)\neq 0\qquad \qquad Sub-Case\text{ }d  \nonumber \\
\Delta _{12}+\Delta _{11} &=&(-1)\neq 0\quad And/Or\quad \Delta _{21}+\Delta
_{22}=(-1)\neq 0\qquad \qquad Sub-Case\text{ }e  \nonumber \\
\Delta _{11} &=&(1)\neq 0\quad And/Or\quad \Delta _{21}+\Delta _{22}+\Delta
_{12}=(-3)\neq 0\qquad Sub-Case\text{ }f  \nonumber \\
&&
\end{eqnarray}%
so as there is at least one \ inequality satisfied for each Sub-Case we see
that the requirements for $S\leq 2\ $are satisfied.

For $S\leq 1\ $we have%
\begin{eqnarray}
\Delta _{12} &=&(-2)\neq 0\quad And/Or\quad \Delta _{22}=(1)\neq 0\quad
And/Or\quad \Delta _{11}+\Delta _{21}=(-1)\neq 0\qquad Sub-Case\text{ }a 
\nonumber \\
\Delta _{11}+\Delta _{12} &=&(-1)\neq 0\quad And/Or\quad \Delta
_{22}=(1)\neq 0\quad And/Or\quad \Delta _{21}=(-2)\neq 0\qquad Sub-Case\text{
}b  \nonumber \\
\Delta _{11} &=&(1)\neq 0\quad And/Or\quad \Delta _{22}+\Delta
_{12}=(-1)\neq 0\quad And/Or\quad \Delta _{21}=(-2)\neq 0\qquad Sub-Case%
\text{ }c  \nonumber \\
\Delta _{11} &=&(1)\neq 0\quad And/Or\quad \Delta _{12}=(-2)\neq 0\quad
And/Or\quad \Delta _{22}+\Delta _{21}=(-1)\neq 0\qquad \qquad Sub-Case\text{ 
}d  \nonumber \\
&&
\end{eqnarray}%
so as there is at least one \ inequality satisfied for each Sub-Case we see
that the requirements for $S\leq 1\ $are satisfied.

Thus, the requirements for $S\leq 3$, $S\leq 2$ and $S\leq 1$.are all
satisfied and we can conclude overall that NLHVT predicts $S\leq 1$\ for
this choice of outcome shifts.$\medskip $

\subsection{Quantum Theory Test}

Here we compare the quantum theory expressions for $S$ with those from NLHVT
based on a specic choice for the $\Delta _{ij}$.

In QTHY\ the result for the quantity $S$ is given by 
\begin{eqnarray}
S &=&\dsum\limits_{jl}P(\alpha _{j},\beta _{l}|\Omega _{A1},\Omega
_{B1},\rho )_{Q}\;\delta _{j,l+\Delta _{11}}+\dsum\limits_{kl}P(\alpha
_{k},\beta _{l}|\Omega _{A2},\Omega _{B1},\rho )_{Q}\;\delta _{l,k+\Delta
_{12}}  \nonumber \\
&&+\dsum\limits_{km}P(\alpha _{k},\beta _{m}|\Omega _{A2},\Omega _{B2},\rho
)_{Q}\;\delta _{k,m+\Delta _{22}}+\dsum\limits_{jm}P(\alpha _{j},\beta
_{m}|\Omega _{A1},\Omega _{B2},\rho )_{Q}\;\delta _{m,j+\Delta _{21}} 
\nonumber \\
&&
\end{eqnarray}%
and using the QTHY\ result (\ref{Eq.QThyResultJointProb}) for $P(\alpha
_{k},\beta _{l}|\Omega _{Aa},\Omega _{Bb},\rho )_{Q}$ in the case of the 
\emph{maximally entangled state} we obtain%
\begin{eqnarray}
S &=&\frac{1}{2d^{3}}\;\dsum\limits_{jl}\frac{1}{\sin ^{2}(\frac{\pi }{d}%
(j-l+\theta _{1}+\phi _{1}))}\;\delta _{j,l+\Delta _{11}}  \nonumber \\
&&+\frac{1}{2d^{3}}\;\dsum\limits_{kl}\frac{1}{\sin ^{2}(\frac{\pi }{d}%
(k-l+\theta _{2}+\phi _{1}))}\;\delta _{l,k+\Delta _{12}}  \nonumber \\
&&+\frac{1}{2d^{3}}\;\dsum\limits_{km}\frac{1}{\sin ^{2}(\frac{\pi }{d}%
(k-m+\theta _{2}+\phi _{2}))}\;\delta _{k,m+\Delta _{22}}  \nonumber \\
&&+\frac{1}{2d^{3}}\;\dsum\limits_{jm}\frac{1}{\sin ^{2}(\frac{\pi }{d}%
(j-m+\theta _{1}+\phi _{2}))}\;\delta _{m,j+\Delta _{21}}  \nonumber \\
&=&\frac{1}{2d^{3}}\;\dsum\limits_{l}\frac{1}{\sin ^{2}(\frac{\pi }{d}%
(\Delta _{11}+\theta _{1}+\phi _{1}))}\;+\frac{1}{2d^{3}}\;\dsum\limits_{l}%
\frac{1}{\sin ^{2}(\frac{\pi }{d}(-\Delta _{12}+\theta _{2}+\phi _{1}))}\; 
\nonumber \\
&&+\frac{1}{2d^{3}}\;\dsum\limits_{m}\frac{1}{\sin ^{2}(\frac{\pi }{d}%
(\Delta _{22}+\theta _{2}+\phi _{2}))}\;+\frac{1}{2d^{3}}\;\dsum\limits_{m}%
\frac{1}{\sin ^{2}(\frac{\pi }{d}(-\Delta _{21}+\theta _{1}+\phi _{2}))}\; 
\nonumber \\
&=&\frac{1}{2d^{2}}\;\frac{1}{\sin ^{2}(\frac{\pi }{d}(\Delta _{11}+\theta
_{1}+\phi _{1}))}\;+\frac{1}{2d^{2}}\;\frac{1}{\sin ^{2}(\frac{\pi }{d}%
(-\Delta _{12}+\theta _{2}+\phi _{1}))}  \nonumber \\
&&+\frac{1}{2d^{2}}\;\frac{1}{\sin ^{2}(\frac{\pi }{d}(\Delta _{22}+\theta
_{2}+\phi _{2}))}\;+\frac{1}{2d^{2}}\;\frac{1}{\sin ^{2}(\frac{\pi }{d}%
(-\Delta _{21}+\theta _{1}+\phi _{2}))}  \nonumber \\
&&  \label{Eq.QThyResultS_CGLMP}
\end{eqnarray}%
for $S$ in the\textbf{\ }\emph{QTHY approach}. Note that adding an integer
multiple of $d$ to the $\Delta _{ij}$ does not alter the result.

Note that although the four sets of measurements for the result for $S$
involve\textbf{\ }\emph{one pair}\textbf{\ }of the two sub-system
observables at a time - ($\Omega _{A1},\Omega _{B1}$), ($\Omega _{A1},\Omega
_{B1}$), ($\Omega _{A1},\Omega _{B1}$) or \ ($\Omega _{A1},\Omega _{B1}$)
the outcomes for the\textbf{\ }\emph{other pair}\textbf{\emph{\ }}of
observables are\textbf{\ }\emph{disregarded. }In an earlier paper \cite%
{Dalton21a} we pointed out that although for HVT (which involves non-quantum
observables) the way the other pair of observable outcomes was disregared
was unambiguous, this was\textbf{\ }\emph{not}\textbf{\ }the case in quantum
theory. Options included (a) not measuring the other pair \ observables at
all (b) measuring the other pair of observables and then discarding the
results - and this could be done in\textbf{\ }\emph{two ways}\textbf{\ }%
depending on whether the disregarded observables were measured\textbf{\ }%
\emph{first or}\textbf{\emph{\ }}\emph{second.}\textbf{\ }The quantum
expressions differ for (a) and (b) since the latter involve a projection
operation into the space of the disregarded observables. Any of these would
be a possible quantum prediction for comparison purposes. However, as it is
only necessary to show HVT is inconsistent with\emph{\ one} possible quantum
prediction we choose the simplest option, namely that in each term the other
observables are\textbf{\ }\emph{not measured }at all. \ This was the choice
made by CGLMP and Eq (\ref{Eq.QThyResultJointProb}) for the quantum theory
expression $P(\alpha _{k},\beta _{l}|\Omega _{Aa},\Omega _{Bb},\rho )_{Q}$
is determined this way.

We could choose as a special case $\Delta _{ij}$ in the form $\Delta
_{ij}=d\times \Delta _{ij}^{\ast }$ where the $\Delta _{ij}^{\ast }$ are all
integers. In this case, using $\sin ^{2}(\pi \Delta _{ij}^{\ast }+\eta
)=(\sin (\pi \Delta _{ij}^{\ast })\cos (\eta )+\cos (\pi \Delta _{ij}^{\ast
})\sin (\eta ))^{2}=\sin ^{2}(\eta )$ in Eq (\ref{Eq.QThyResultS_CGLMP}), we
obtain for the\textbf{\ }\emph{QTHY approach}%
\begin{eqnarray}
S &=&\frac{1}{2d^{2}}\;\frac{1}{\sin ^{2}(\frac{\pi }{d}(d\Delta _{11}^{\ast
}+\frac{1}{4}))}\;+\frac{1}{2d^{2}}\;\frac{1}{\sin ^{2}(\frac{\pi }{d}%
(-d\Delta _{12}^{\ast }+\frac{3}{4}))}  \nonumber \\
&&+\frac{1}{2d^{2}}\;\frac{1}{\sin ^{2}(\frac{\pi }{d}(d\Delta _{22}^{\ast }+%
\frac{1}{4}))}\;+\frac{1}{2d^{2}}\;\frac{1}{\sin ^{2}(\frac{\pi }{d}%
(-d\Delta _{21}^{\ast }-\frac{1}{4}))}  \nonumber \\
&=&\frac{1}{2d^{2}}\;\left( \frac{1}{\sin ^{2}(\frac{\pi }{4d})}+\frac{1}{%
\sin ^{2}(\frac{3\pi }{4d})}+\frac{1}{\sin ^{2}(\frac{\pi }{4d})}+\frac{1}{%
\sin ^{2}(\frac{\pi }{4d})}\right)  \nonumber \\
&=&\frac{1}{2d^{2}}\;\left( \frac{3}{\sin ^{2}(\frac{\pi }{4d})}+\frac{1}{%
\sin ^{2}(\frac{3\pi }{4d})}\right)  \label{Eq.QThyResultforSSpecialDeltas}
\end{eqnarray}%
We have used the values $a=1,2\quad \theta _{1}=0,\theta _{2}=1/2$ and $%
b=1,2\quad \phi _{1}=1/4,\phi _{2}=-1/4$ from Eqs (\ref%
{Eq.EigenvectorsObserA}), (\ref{Eq.EigenvectorsObservB}) for \ the
eigenvectors\textbf{. }This last result shows that in quantum theory the
value of $S$ is \emph{independent} of the $\Delta _{ij}^{\ast }$. It only
depends on the integer $d$. The NLHVT\ Bell inequalities for $S$ do of
course depend on the $\Delta _{ij}^{\ast }$.

For the choice $d=10$ we find that in QTHY $S\approx 2.52$, so this exceeds $%
S=2$ and $S=1$ though not $S=4$ and $S=3$ that could occur in NLHVT
depending on the requirements for these cases being met. For the choice $%
(\Delta _{11}^{\ast },\Delta _{12}^{\ast },\Delta _{22}^{\ast },\Delta
_{21}^{\ast })=(1,-2,1,-2)$ it is easily confirmed that the requirements for 
$S\leq 4$ are \emph{not }met. However the requirements for $S\leq 3$, for $%
S\leq 2$\ and for $S\leq 1$\ are\textbf{\emph{\ all} }met. . Other choices
are also be made. For example with $(\Delta _{11}^{\ast },\Delta _{12}^{\ast
},\Delta _{22}^{\ast },\Delta _{21}^{\ast })=(1,-3,1,-3)$ it is easily
confirmed that the requirements for $S\leq 4$ are \emph{not} met. However
the requirements for $S\leq 3,S\leq 2$ and $S\leq 1$ are \emph{all} met. In
\ both these \ situations NLHVT predicts $S\leq 1$\ - this being the
smallest of the three inequalities. Thus we have found cases for $d=10$
where the \emph{quantum theory} prediction that $S=2.52$ is \emph{inconsisten%
}t with the \emph{NLHVT Bell inequality} prediction of $S\leq 1.$

We could also choose $(\Delta _{11},\Delta _{12},\Delta _{22},\Delta
_{21})=(0,1,0,0)$ as in the original CGLMP Bell inequality (\ref%
{Eq.CHSHBelllneq1}). For this choice the requirements for $S\leq 3$, for $%
S\leq 2$\ and for $S\leq 1$\ are\textbf{\emph{\ all} }met\textbf{. }.In this
case the CGLMP\ Bell inequality predicts $S\leq 1$. For this choice the QTHY%
\textbf{\ }expression for $S$ is given by 
\begin{eqnarray}
S &=&\frac{1}{2d^{2}}\;\frac{1}{\sin ^{2}(\frac{\pi }{d}(\frac{1}{4}))}\;+%
\frac{1}{2d^{2}}\;\frac{1}{\sin ^{2}(\frac{\pi }{d}(-1+\frac{3}{4}))} 
\nonumber \\
&&+\frac{1}{2d^{2}}\;\frac{1}{\sin ^{2}(\frac{\pi }{d}(\frac{1}{4}))}\;+%
\frac{1}{2d^{2}}\;\frac{1}{\sin ^{2}(\frac{\pi }{d}(-\frac{1}{4}))} 
\nonumber \\
&=&\frac{2}{d^{2}}\frac{1}{\sin ^{2}(\frac{\pi }{4d})}
\label{Eq.QThyResultforS0100Case}
\end{eqnarray}

For the case of $d=2$ this gives $S=3.41$. For the choice $(\Delta
_{11},\Delta _{12},\Delta _{22},\Delta _{21})=(0,1,0,0)$ the requirements
for $S\leq 4$ are \emph{not }met, but those for\textbf{\ }$S\leq 1$\textbf{\ 
}are met\textbf{,}.However NLHVT is inconsistent with QTHY in this choice of 
$(\Delta _{11},\Delta _{12},\Delta _{22},\Delta _{21})$ - as it was for LHVT
where in both \ types of HVT the inequality $S\leq 1$\ is violated in Q
THY.\medskip

\section{SUMMARY\ AND\ CONCLUSION}

A brief review of the origins of local hidden variable theory, Bell local
states and Bell inequalities has been presented. We have also outlined the
basic justification for LHVT. The classification of Bell local states into
three categorise based on the\emph{\ local hidden quantum state }approach of
Wiseman \cite{Wiseman07a} has been explained. We outlined how the Bell
inequalities provide tests for whether local hidden variable theory is in
conflict with quantum theory and also which categories of Bell local states
exhbit quantum entanglement or EPR steering.

The new result in this paper is that a Bell-type \emph{non-local} hidden
variable theory has been presented, and Bell inequalities obtained based on
this NLHVT. This is based on the approach of Collins, Gisin, Linden, Massar
Popescu (CGLMP)\ for\emph{\ bipartite }systems, where the basic joint
probability involves the outcomes for two observables for each\textbf{\ \ }%
sub-system \ In addition to showing that the well-known CHSH inequality can
also be proved \ for such a NLHVT, we have also considered CGLMP\ Bell
inequalities involving the quantity $S$ that involves joint probabilities
for outcomes from a\textbf{\ }\emph{pair}\textbf{\emph{\ }}of sub-system
observables, \ \emph{one} in each sub-system.\textbf{\ }\emph{Sufficiency}
conditions for the \emph{outcome shifts} have been obtained for CGLMP
inequalities $S\leq 4,S\leq 3,S\leq 2$ and $S\leq 1$. It has been shown that
in a \emph{maximally entangled} state in a bipartite system clear \emph{%
quantum theory }violations of these CGLMP\ Bell inequalities occur , thus
showing that it is \emph{not posible} to underpin or complete quantum theory
via a CGLMP\ \ Bell-type non-local hidden variable theory. The same
sufficiency conditions also show that there is \emph{also} a quantum theory
violation for the CGLMP version of LHVT.\textbf{\ }

It should be pointed out that the observables chosen have \emph{no obvious}
physical meaning. Also the experimental measurements for the NLHVT Bell
inequality terms would \emph{difficult to perform.} Nor would the maximally
entangled state be \emph{easy to prepare}. However these issues are \emph{%
irrelevant} as all that is needed is that the observables are \emph{possible}
quantum observables, the state is a \emph{possible} quantum state and the
Bell inequality measurements \emph{could} be carried out.

The result that a Bell-type non-local hidden variable theory is also in
conflict with quantum theory is of \emph{some significance} since it shows
that quantum theory \emph{can not} be understood \ in terms of Bell-type
hidden variable theory. This is \emph{irrespective} of whether the hidden
variable theory is \emph{local} or\emph{\ non-local}. \medskip

\section{ACKNOWLEDGEMENTS}

The author wishes to acknowledge discussions on quantun foundations with
several colleagues - M Reid, S Barnett, B Garraway, L Heaney, J Goold, T
Busch, L Rosales, R Teh, B Opanchuk and P Drummond.\bigskip \pagebreak

\section{APPENDIX A - LOCAL\ HIDDEN\ QUANTUM STATES, ENTANGLEMENT\ \& EPR
STEERING}

Quantum states can be divided into \emph{entangled} and \emph{separable}
states, but in \emph{LHVT} can also be divided into \emph{Bell local} and 
\emph{Bell non-local} states. These categories \emph{overlap }\cite%
{Werner89a}.\smallskip

\subsection{Separable States}

For \emph{separable} states, prepare \emph{sub-systems} $A$, $B$ in \emph{%
quantum states} $\widehat{\rho }_{R}^{A}$, $\widehat{\rho }_{R}^{B}$, where 
\emph{combined} state $\widehat{\rho }_{R}^{A}\otimes \widehat{\rho }%
_{R}^{B} $ occurs with \emph{classical} \emph{probability} $P_{R}$ \cite%
{Werner89a}. With density opr $\widehat{\rho }_{sep}=\sum_{R}P_{R}\;\widehat{%
\rho }_{R}^{A}\otimes \widehat{\rho }_{R}^{B}$ and $P(\alpha |\Omega
_{A},R)=Tr_{A}(\widehat{\Pi }_{\alpha }^{A}\widehat{\rho }_{R}^{A})$
etc.,joint outcome prob is given by a \emph{LHVT form}: 
\begin{equation}
P(\alpha ,\beta |\Omega _{A},\Omega _{B},\rho
_{sep})_{Q}=\sum_{R}P_{R}\,P(\alpha |\Omega _{A},R)\times P(\beta |\Omega
_{B},R)
\end{equation}%
This is of the \emph{same form} as for Bell local states, showing that all 
\emph{separable} states are \emph{Bell-local }($R\rightarrow \lambda $).
Hence \emph{Bell non-local} states must be \emph{entangled}.

However, \emph{some} Bell local states are also \emph{entangled} \emph{\ }%
\cite{Werner89a}. For \emph{Bell-local} states which are \emph{separable}, 
\emph{both} sub-systems are associated with a \emph{quantum} density
operator. Hence idea of a \emph{local hidden quantum state }(\emph{LHQS})\
was introduced \cite{Wiseman07a}, \cite{Jones07a}, \cite{Cavalcanti09a}\emph{%
.}\smallskip

\subsection{Local Hidden Quantum State}

Bell-local states in bipartite systems can be further divided into \emph{%
three} disjoint categories, depending on \emph{two}, \emph{one} or \emph{none%
} of the sub-systems being associated with a \emph{local hidden quantum
state }\cite{Wiseman07a}, \cite{Jones07a}, \cite{Cavalcanti09a}. LHQS $%
\widehat{\rho }_{C}(\lambda )$ are to be \emph{possible} quantum states $-$
these must comply with conditions such as the \emph{super-selection rules}
of there being \emph{no coherences} between states with \emph{differing}
(massive) particle numbers. \emph{\smallskip }

\subsection{Bell-Local States - Three Categories}

For the \emph{bipartite} case - there are \emph{three} categories of \emph{%
Bell-local} states%
\begin{eqnarray}
P(\alpha ,\beta |\Omega _{A},\Omega _{B},c)_{LHVT} &=&\tsum\limits_{\lambda
}P_{Q}(\alpha |\Omega _{A},\lambda )\,P_{Q}(\beta |\Omega _{B},\lambda
)\,P(\lambda |c)\qquad Cat\;1  \nonumber \\
P(\alpha ,\beta |\Omega _{A},\Omega _{B},c)_{LHVT} &=&\tsum\limits_{\lambda
}P(\alpha |\Omega _{A},\lambda )\,P_{Q}(\beta |\Omega _{B},\lambda
)\,P(\lambda |c)\qquad Cat\;2  \nonumber \\
P(\alpha ,\beta |\Omega _{A},\Omega _{B},c)_{LHVT} &=&\tsum\limits_{\lambda
}P(\alpha |\Omega _{A},\lambda )\,P(\beta |\Omega _{B},\lambda )\,P(\lambda
|c)\qquad Cat\;3  \nonumber \\
&&
\end{eqnarray}%
where subscript $Q$ indicates sub-system measurement probability determined
from a \emph{local hidden quantum state} $\widehat{\rho }_{C}(\lambda )$,
via $P_{Q}(\gamma |\Omega _{C},\lambda )=Tr_{C}(\widehat{\Pi }_{\gamma }^{C}%
\widehat{\rho }_{C}(\lambda ))${\Large . }Bell inequalities will \emph{differ%
} depending on whether there are sub-systems associated with a LHQS. {\Large %
\medskip }

\subsection{Tests for Quantum Entanglement and EPR Steering{\protect\Large \ 
}}

For the case of \emph{two modes}. we introduce \emph{spin operators} $S_{x}=%
\frac{1}{2}(b^{\dag }a+a^{\dag }b),S_{y}=\frac{1}{2i}(b^{\dag }a-a^{\dag
}b),S_{z}=\frac{1}{2}(b^{\dag }b-a^{\dag }a)$.

\emph{Category 1} states are \emph{same} as \emph{separable} states since 
\emph{both} sub-system probabilities are given by quantum expressions.
Violation of \emph{inequalities} based on Category 1 states $\rightarrow $ 
\emph{Quantum entanglement} (non-separability) occurs. Example: \emph{Spin
squeezing} test for \emph{two mode} entanglement \cite{Dalton14a}{\small \ }$%
\left\langle \Delta S_{x}^{2}\right\rangle <\frac{1}{2}|\left\langle
S_{z}\right\rangle $ or $|\left\langle \Delta S_{y}^{2}\right\rangle <\frac{1%
}{2}|\left\langle S_{z}\right\rangle |$. \ 

For \emph{Category 2} states only \emph{one} sub-system probability ($B$) is
given by quantum expression involving \emph{LHS}\ quantum density operator $%
- $ $\widehat{\rho }^{B}(\lambda )$. Here no EPR steering of $B$ by $A$
occurs $\rightarrow $ Outcomes $\beta $ for $\Omega _{B}$ depend \emph{only}
on $\widehat{\rho }_{B}(\lambda )$ $-$ \emph{not} on $\alpha ,\Omega _{A}$.
Violation of \emph{inequalities} based on Category 2 states $\rightarrow $ 
\emph{EPR steering }of $B$ outcomes by $A$ measurement occurs. Example:
generalised Hillary-Zubairy \emph{planar spin variance} test for \emph{two
mode} EPR steering \cite{Hillary06a}{\small , }\cite{Dalton20a}. $%
\left\langle \Delta S_{x}^{2}\right\rangle +\left\langle \Delta
S_{y}^{2}\right\rangle <(\frac{1}{4}\left\langle N\right\rangle -\frac{1}{2}%
\left\langle S_{z}\right\rangle )$.\bigskip \pagebreak

\subsection{Tests for Bell Non-Locality}

For \emph{Category 3} states \emph{neither} of sub-system probabilities
given by quantum expression - no \emph{local hidden quantum state. }. \emph{%
Inequality} violations based on Category 3 states $\rightarrow $ \emph{Bell
locality cannot }apply\emph{. }Example: \emph{CHSH Bell Inequality (above) }%
for \emph{two spin 1/2} systems. Thus \emph{Bell non-locality is }shown by
violations of inequalities based on

\begin{equation}
P(\alpha ,\beta |\Omega _{A},\Omega _{B},c)_{LHVT}=\dsum\limits_{\lambda
}P(\lambda |c)P(\alpha |\Omega _{A},\lambda )P(\beta |\Omega _{B},\lambda )
\end{equation}

\subsection{Bell - Non-Local States}

A \emph{fourth} category of states are the \emph{Bell non-local} states.

\begin{equation}
P(\alpha ,\beta |\Omega _{A},\Omega _{B},c)_{NLHVT}=\tsum\limits_{\lambda
}P(\alpha ,\beta |\Omega _{A},\Omega _{B},\lambda )\,\,P(\lambda |c)\qquad
Cat\;4
\end{equation}%
Here there is no \emph{separate} $P(\gamma |\Omega _{C},\lambda )\,$for each
sub-system. Violation of inequalities based on Category 4 states show that 
\emph{non-local hidden variable theories }also\emph{\ fail.}

\pagebreak

\subsection{Overall Scheme}

The \emph{overall scheme} can be pictured as (see \cite{Dalton19a}, \cite%
{Dalton20a}{\small ) \ }Here we have \emph{assumed} that Quantum Theory is
underpinned by Hidden Variable Theory - local or non-local and that Bell
non-local states are quantum entangled states. The issue of whether or not
quantum theory is underpinned by a non-local hidden variable theory is the
focus of the main part of this paper.\textbf{\ }\bigskip

\FRAME{ftbpFU}{4.8066in}{3.6123in}{0pt}{\Qcb{{The classifcation of hidden
variable Bell local states into three categories, plus \ the hidden variable
Bell non-local states. Also the classification of quantum states as
separable and entangled. The features of the various types of states are
stated.}}}{}{classificationschemesmk2.jpg}{\special{language "Scientific
Word";type "GRAPHIC";maintain-aspect-ratio TRUE;display "USEDEF";valid_file
"F";width 4.8066in;height 3.6123in;depth 0pt;original-width
9.9998in;original-height 7.4996in;cropleft "0";croptop "1";cropright
"1";cropbottom "0";filename 'ClassificationSchemesMk2.jpg';file-properties
"XNPEU";}}\medskip

\pagebreak

\section{APPENDIX B - REQUIREMENTS\ FOR\ BELL\ TESTS}

In this Appendix \ we will derive the requirements for Bell Inequality tests
in the NLHVT situation.

\subsection{Case 1 - Bell Inequality - All Conditions T}

If $C1=T,C2=T,C3=T,C4=T$ we then have 
\begin{equation}
\left[ 
\begin{array}{cccc}
1 & 0 & -1 & 0 \\ 
0 & -1 & 1 & 0 \\ 
0 & 1 & 0 & -1 \\ 
-1 & 0 & 0 & 1%
\end{array}%
\right] \;\left[ 
\begin{array}{c}
j \\ 
k \\ 
l \\ 
m%
\end{array}%
\right] =\left[ 
\begin{array}{c}
\Delta _{11} \\ 
\Delta _{12} \\ 
\Delta _{22} \\ 
\Delta _{21}%
\end{array}%
\right]
\end{equation}

As $|M|=0$ there is only a non-zero solution for $\mathbf{v}$ if the
following \emph{requirements} are satisfied 
\begin{equation}
\Delta _{11}=0\quad \Delta _{12}=0\quad \Delta _{22}=0\quad \Delta _{21}=0
\end{equation}%
This solution is $j=k=l=m$ and here we have $\delta =4$ \ 

Hence applying this result for\textbf{\ }$\delta $\textbf{\ }in Eq. (\ref%
{Eq.SforNLHVTNew}) 
\begin{eqnarray}
S &\leq &\dsum\limits_{\lambda }P(\lambda |c)\left\{
\dsum\limits_{j,k,l,m}P(\alpha _{j},\alpha _{k},\beta _{l},\beta _{m}|\Omega
_{A1},\Omega _{A2},\Omega _{B1},\Omega _{B2};\lambda )\left( 4\right)
\,\right\}  \nonumber \\
&\leq &\dsum\limits_{\lambda }P(\lambda |c)\;(4)  \nonumber \\
&\leq &4
\end{eqnarray}%
The Bell inequality is then $S\leq 4$. \medskip

\subsection{Case 2 - Bell Inequality - Three Conditions T, One F}

Here there are\emph{\ fou}r sub-cases. In all of thes sub-cases we have%
\textbf{\ }$\delta =3.$

\subsubsection{Sub-Case a - $C1=T,C2=T,C3=T,C4=F$}

Substituting $C1,C2,C3$ into $C4$ we have 
\begin{eqnarray}
m &=&j+\Delta _{21}  \nonumber \\
&=&l+\Delta _{11}+\Delta _{21}  \nonumber \\
&=&k+\Delta _{12}+\Delta _{11}+\Delta _{21}  \nonumber \\
&=&m+\Delta _{22}+\Delta _{12}+\Delta _{11}+\Delta _{21}
\end{eqnarray}

As $C4=F$ it follows that the last equation is false. Hence the following
requirement must be satisfied%
\begin{equation}
\Delta _{22}+\Delta _{12}+\Delta _{11}+\Delta _{21}\neq 0
\end{equation}%
\smallskip

\subsubsection{Sub-Case b - $C1=T,C2=T,C3=F,C4=T$}

Substituting $C1,C2,C4$ into $C3$ we have 
\begin{eqnarray}
k &=&m+\Delta _{22}  \nonumber \\
&=&j+\Delta _{21}+\Delta _{22}  \nonumber \\
&=&l+\Delta _{11}+\Delta _{21}+\Delta _{22}  \nonumber \\
&=&k+\Delta _{12}+\Delta _{11}+\Delta _{21}+\Delta _{22}
\end{eqnarray}

As $C3=F$ it follows that the last equation is false. Hence the following 
\emph{requirement} must be satisfied%
\begin{equation}
\Delta _{12}+\Delta _{11}+\Delta _{21}+\Delta _{22}\neq 0
\end{equation}%
\smallskip

\subsubsection{Sub-Case c - $C1=T,C2=F,C3=T,C4=T$}

The requirement and the Bell inequality is the same as for Sub-Cases a and
b. \smallskip

\subsubsection{Sub-Case d - $C1=F,C2=T,C3=T,C4=T$}

The requirement and the Bell inequality is the same as for Sub-Cases a, b
and c. \smallskip

\subsubsection{All Sub-Cases of Case 2}

Overall, we see that if the requirement is satisfied, we can say $\delta =3.$

Hence applying this result for\textbf{\ }$\delta $ in Eq. (\ref%
{Eq.SforNLHVTNew}) 
\begin{eqnarray}
S &\leq &\dsum\limits_{\lambda }P(\lambda |c)\left\{
\dsum\limits_{j,k,l,m}P(\alpha _{j},\alpha _{k},\beta _{l},\beta _{m}|\Omega
_{A1},\Omega _{A2},\Omega _{B1},\Omega _{B2};\lambda )\left( 3\right)
\,\right\}  \nonumber \\
&\leq &\dsum\limits_{\lambda }P(\lambda |c)\;(3)  \nonumber \\
&\leq &3
\end{eqnarray}

From above the following \emph{requirement} must be satisfied for Case 2 to
cover all sub-cases. 
\begin{equation}
\Delta _{12}+\Delta _{11}+\Delta _{21}+\Delta _{22}\neq 0
\end{equation}%
The Bell inequality is then $S\leq 3$. \medskip \bigskip

\subsection{Case 3 - Bell Inequality - Two Conditions T, Two Conditions F}

Here there are\emph{\ six} sub-cases. In all of these sub-cases $\delta =2$.

\subsubsection{Sub-Case a $C1=T,C2=T,C3=F,C4=F$}

Substituting $C1,C2$ into $C3,$ $C4$ we have%
\begin{eqnarray}
k &=&m+\Delta _{22}  \nonumber \\
m &=&j+\Delta _{21}  \nonumber \\
&=&l+\Delta _{11}+\Delta _{21}  \nonumber \\
&=&k+\Delta _{12}+\Delta _{11}+\Delta _{21}
\end{eqnarray}

Thus we have the marix equation 
\begin{equation}
\left[ 
\begin{array}{cc}
+1 & -1 \\ 
-1 & +1%
\end{array}%
\right] \;\left[ 
\begin{array}{c}
k \\ 
m%
\end{array}%
\right] =\left[ 
\begin{array}{c}
\Delta _{22} \\ 
\Delta _{12}+\Delta _{11}+\Delta _{21}%
\end{array}%
\right]
\end{equation}%
Since $C3=F,C4=F$ we require there\emph{\ no}t to be a solution.

As $|M|=0$ there is \emph{not} a non-zero solution for $\mathbf{v}$ if the
following \emph{requirements} are satisfied%
\begin{equation}
\Delta _{22}\neq 0\quad And/Or\quad \Delta _{12}+\Delta _{11}+\Delta
_{21}\neq 0
\end{equation}%
\smallskip

\subsubsection{Sub-Case b $C1=T,C2=F,C3=T,C4=F$\protect\bigskip}

Substituting $C1,C3$ into $C2,$ $C4$ we have%
\begin{eqnarray}
l &=&k+\Delta _{12}  \nonumber \\
&=&m+\Delta _{22}+\Delta _{12}  \nonumber \\
m &=&j+\Delta _{21}  \nonumber \\
&=&l+\Delta _{11}+\Delta _{21}
\end{eqnarray}

Thus we have the marix equation 
\begin{equation}
\left[ 
\begin{array}{cc}
+1 & -1 \\ 
-1 & +1%
\end{array}%
\right] \;\left[ 
\begin{array}{c}
l \\ 
m%
\end{array}%
\right] =\left[ 
\begin{array}{c}
\Delta _{22}+\Delta _{12} \\ 
\Delta _{11}+\Delta _{21}%
\end{array}%
\right]
\end{equation}%
Since $C2=F,C4=F$ we require there\emph{\ no}t to be a solution.

As $|M|=0$ there is \emph{not} a non-zero solution for $\mathbf{v}$ if the
following \emph{requirements} are satisfied%
\begin{equation}
\Delta _{22}+\Delta _{12}\neq 0\quad And/Or\quad \Delta _{11}+\Delta
_{21}\neq 0
\end{equation}%
\smallskip

\subsubsection{Sub-Case c $C1=T,C2=F,C3=F,C4=T$}

Substituting $C1,C4$ into $C2,$ $C3$ we have%
\begin{eqnarray}
l &=&k+\Delta _{12}  \nonumber \\
k &=&m+\Delta _{22}  \nonumber \\
&=&j+\Delta _{21}+\Delta _{22}  \nonumber \\
&=&l+\Delta _{11}+\Delta _{21}+\Delta _{22}
\end{eqnarray}

Thus we have the marix equation 
\begin{equation}
\left[ 
\begin{array}{cc}
-1 & +1 \\ 
+1 & -1%
\end{array}%
\right] \;\left[ 
\begin{array}{c}
k \\ 
l%
\end{array}%
\right] =\left[ 
\begin{array}{c}
\Delta _{12} \\ 
\Delta _{11}+\Delta _{21}+\Delta _{22}%
\end{array}%
\right]
\end{equation}%
Since $C2=F,C3=F$ we require there\emph{\ no}t to be a solution.

As $|M|=0$ there is \emph{not} a non-zero solution for $\mathbf{v}$ if the
following \emph{requirements} are satisfied%
\begin{equation}
\Delta _{12}\neq 0\quad And/Or\quad \Delta _{11}+\Delta _{21}+\Delta
_{22}\neq 0
\end{equation}%
\smallskip

\subsubsection{Sub-Case d $C1=F,C2=T,C3=T,C4=F$}

Substituting $C2,C3$ into $C1,$ $C4$ we have%
\begin{eqnarray}
j &=&l+\Delta _{11}  \nonumber \\
&=&k+\Delta _{12}+\Delta _{11}  \nonumber \\
&=&m+\Delta _{22}+\Delta _{12}+\Delta _{11}  \nonumber \\
m &=&j+\Delta _{21}
\end{eqnarray}

Thus we have the marix equation 
\begin{equation}
\left[ 
\begin{array}{cc}
+1 & -1 \\ 
-1 & +1%
\end{array}%
\right] \;\left[ 
\begin{array}{c}
j \\ 
m%
\end{array}%
\right] =\left[ 
\begin{array}{c}
\Delta _{22}+\Delta _{12}+\Delta _{11} \\ 
\Delta _{21}%
\end{array}%
\right]
\end{equation}%
Since $C1=F,C4=F$ we require there\emph{\ no}t to be a solution.

As $|M|=0$ there is \emph{not} a non-zero solution for $\mathbf{v}$ if the
following \emph{requirements} are satisfied%
\begin{equation}
\Delta _{22}+\Delta _{12}+\Delta _{11}\neq 0\quad And/Or\quad \Delta
_{21}\neq 0
\end{equation}%
\smallskip

\subsubsection{Sub-Case e $C1=F,C2=T,C3=F,C4=T$}

Substituting $C2,C4$ into $C1,$ $C3$ we have%
\begin{eqnarray}
j &=&l+\Delta _{11}  \nonumber \\
&=&k+\Delta _{12}+\Delta _{11}  \nonumber \\
k &=&m+\Delta _{22}  \nonumber \\
&=&j+\Delta _{21}+\Delta _{22}
\end{eqnarray}

Thus we have the marix equation 
\begin{equation}
\left[ 
\begin{array}{cc}
+1 & -1 \\ 
-1 & +1%
\end{array}%
\right] \;\left[ 
\begin{array}{c}
j \\ 
k%
\end{array}%
\right] =\left[ 
\begin{array}{c}
\Delta _{12}+\Delta _{11} \\ 
\Delta _{21}+\Delta _{22}%
\end{array}%
\right]
\end{equation}%
Since $C1=F,C3=F$ we require there\emph{\ no}t to be a solution.

As $|M|=0$ there is \emph{not} a non-zero solution for $\mathbf{v}$ if the
following \emph{requirements} are satisfied%
\begin{equation}
\Delta _{12}+\Delta _{11}\neq 0\quad And/Or\quad \Delta _{21}+\Delta
_{22}\neq 0
\end{equation}%
\smallskip

\subsubsection{Sub-Case f $C1=F,C2=F,C3=T,C4=T$}

Substituting $C3,C4$ into $C1,$ $C2$ we have%
\begin{eqnarray}
j &=&l+\Delta _{11}  \nonumber \\
l &=&k+\Delta _{12}  \nonumber \\
&=&m+\Delta _{22}+\Delta _{12}  \nonumber \\
&=&j+\Delta _{21}+\Delta _{22}+\Delta _{12}
\end{eqnarray}

Thus we have the marix equation 
\begin{equation}
\left[ 
\begin{array}{cc}
-1 & +1 \\ 
-1 & +1%
\end{array}%
\right] \;\left[ 
\begin{array}{c}
l \\ 
j%
\end{array}%
\right] =\left[ 
\begin{array}{c}
\Delta _{11} \\ 
\Delta _{21}+\Delta _{22}+\Delta _{12}%
\end{array}%
\right]
\end{equation}%
Since $C1=F,C2=F$ we require there\emph{\ no}t to be a solution.

As $|M|=0$ there is \emph{not} a non-zero solution for $\mathbf{v}$ if the
following \emph{requirements} are satisfied%
\begin{equation}
\Delta _{11}\neq 0\quad And/Or\quad \Delta _{21}+\Delta _{22}+\Delta
_{12}\neq 0
\end{equation}%
\smallskip

\subsubsection{All Sub-Cases of Case 3}

Overall, we see that if the requirement is satisfied, we can say $\delta =2$.

Hence applying this value of $\delta $ in Eq. (\ref{Eq.SforNLHVTNew})%
\begin{eqnarray}
S &\leq &\dsum\limits_{\lambda }P(\lambda |c)\left\{
\dsum\limits_{j,k,l,m}P(\alpha _{j},\alpha _{k},\beta _{l},\beta _{m}|\Omega
_{A1},\Omega _{A2},\Omega _{B1},\Omega _{B2};\lambda )\left( 2\right)
\,\right\}  \nonumber \\
&\leq &\dsum\limits_{\lambda }P(\lambda |c)\;(2)  \nonumber \\
&\leq &2
\end{eqnarray}

From above the following \emph{requirements} must be satisfied for Case 3 to
cover all sub-cases.%
\begin{eqnarray}
\Delta _{22} &\neq &0\quad And/Or\quad \Delta _{12}+\Delta _{11}+\Delta
_{21}\neq 0\qquad Sub-Case\text{ }a  \nonumber \\
\Delta _{22}+\Delta _{12} &\neq &0\quad And/Or\quad \Delta _{11}+\Delta
_{21}\neq 0\qquad \qquad Sub-Case\text{ }b  \nonumber \\
\Delta _{12} &\neq &0\quad And/Or\quad \Delta _{11}+\Delta _{21}+\Delta
_{22}\neq 0\qquad Sub-Case\text{ }c  \nonumber \\
\Delta _{22}+\Delta _{12}+\Delta _{11} &\neq &0\quad And/Or\quad \Delta
_{21}\neq 0\qquad \qquad Sub-Case\text{ }d  \nonumber \\
\Delta _{12}+\Delta _{11} &\neq &0\quad And/Or\quad \Delta _{21}+\Delta
_{22}\neq 0\qquad \qquad Sub-Case\text{ }e  \nonumber \\
\Delta _{11} &\neq &0\quad And/Or\quad \Delta _{21}+\Delta _{22}+\Delta
_{12}\neq 0\qquad Sub-Case\text{ }f  \nonumber \\
&&
\end{eqnarray}%
The Bell inequality is then $S\leq 2$. \medskip

\subsection{Case 4 Bell Inequality - One Condition T, Three Conditions F}

Here there are\emph{\ four} sub-cases. In all these \ sub-cases \ $\delta =1$%
.

\subsubsection{Sub-Case a $C1=T,C2=F,C3=F,C4=F$}

Substituting $C1$ into $C2,$ $C3,$ $C4$ we have%
\begin{eqnarray}
l &=&k+\Delta _{!2}  \nonumber \\
k &=&m+\Delta _{22}  \nonumber \\
m &=&j+\Delta _{21}  \nonumber \\
&=&l+\Delta _{11}+\Delta _{21}
\end{eqnarray}%
Thus we have the marix equation 
\begin{equation}
\left[ 
\begin{array}{ccc}
-1 & +1 & 0 \\ 
+1 & 0 & -1 \\ 
0 & -1 & +1%
\end{array}%
\right] \;\left[ 
\begin{array}{c}
k \\ 
l \\ 
m%
\end{array}%
\right] =\left[ 
\begin{array}{c}
\Delta _{!2} \\ 
\Delta _{22} \\ 
\Delta _{11}+\Delta _{21}%
\end{array}%
\right]
\end{equation}%
Since $C2=F,C3=F,C4=F$ we require there\emph{\ no}t to be a solution.

As $|M|=0$ there is \emph{not} a non-zero solution for $\mathbf{v}$ if the
following \emph{requirements} are satisfied%
\begin{equation}
\Delta _{!2}\neq 0\quad And/Or\quad \Delta _{22}\neq 0\quad And/Or\quad
\Delta _{11}+\Delta _{21}\neq 0
\end{equation}%
\smallskip

\subsubsection{Sub-Case b $C1=F,C2=T,C3=F,C4=F$}

Substituting $C2$ into $C1,$ $C3,$ $C4$ we have%
\begin{eqnarray}
j &=&l+\Delta _{11}  \nonumber \\
&=&k+\Delta _{!2}+\Delta _{11}  \nonumber \\
k &=&m+\Delta _{22}  \nonumber \\
m &=&j+\Delta _{21}
\end{eqnarray}%
Thus we have the marix equation 
\begin{equation}
\left[ 
\begin{array}{ccc}
+1 & -1 & 0 \\ 
0 & +1 & -1 \\ 
-1 & 0 & +1%
\end{array}%
\right] \;\left[ 
\begin{array}{c}
j \\ 
k \\ 
m%
\end{array}%
\right] =\left[ 
\begin{array}{c}
\Delta _{!2}+\Delta _{11} \\ 
\Delta _{22} \\ 
\Delta _{21}%
\end{array}%
\right]
\end{equation}%
Since $C1=F,C3=F,C4=F$ we require there\emph{\ no}t to be a solution.

As $|M|=0$ there is \emph{not} a non-zero solution for $\mathbf{v}$ if the
following \emph{requirements} are satisfied%
\begin{equation}
\Delta _{!2}+\Delta _{11}\neq 0\quad And/Or\quad \Delta _{22}\neq 0\quad
And/Or\quad \Delta _{21}\neq 0
\end{equation}%
\smallskip

\subsubsection{Sub-Case c $C1=F,C2=F,C3=T,C4=F$}

Substituting $C3$ into $C1,$ $C2,$ $C4$ we have%
\begin{eqnarray}
j &=&l+\Delta _{11}  \nonumber \\
l &=&k+\Delta _{12}  \nonumber \\
&=&m+\Delta _{22}+\Delta _{12}  \nonumber \\
m &=&j+\Delta _{21}
\end{eqnarray}%
Thus we have the marix equation 
\begin{equation}
\left[ 
\begin{array}{ccc}
+1 & -1 & 0 \\ 
0 & +1 & -1 \\ 
-1 & 0 & +1%
\end{array}%
\right] \;\left[ 
\begin{array}{c}
j \\ 
l \\ 
m%
\end{array}%
\right] =\left[ 
\begin{array}{c}
\Delta _{11} \\ 
\Delta _{22}+\Delta _{12} \\ 
\Delta _{21}%
\end{array}%
\right]
\end{equation}%
Since $C1=F,C2=F,C4=F$ we require there\emph{\ no}t to be a solution.

As $|M|=0$ there is \emph{not} a non-zero solution for $\mathbf{v}$ if the
following \emph{requirements} are satisfied%
\begin{equation}
\Delta _{11}\neq 0\quad And/Or\quad \Delta _{22}+\Delta _{12}\neq 0\quad
And/Or\quad \Delta _{21}\neq 0
\end{equation}%
\smallskip

\subsubsection{Sub-Case d $C1=F,C2=F,C3=F,C4=T$}

Substituting $C4$ into $C1,$ $C2,$ $C3$ we have%
\begin{eqnarray}
j &=&l+\Delta _{11}  \nonumber \\
l &=&k+\Delta _{12}  \nonumber \\
k &=&m+\Delta _{22}  \nonumber \\
&=&j+\Delta _{21}+\Delta _{22}
\end{eqnarray}%
Thus we have the marix equation 
\begin{equation}
\left[ 
\begin{array}{ccc}
+1 & 0 & -1 \\ 
0 & -1 & +1 \\ 
-1 & +1 & 0%
\end{array}%
\right] \;\left[ 
\begin{array}{c}
j \\ 
k \\ 
l%
\end{array}%
\right] =\left[ 
\begin{array}{c}
\Delta _{11} \\ 
\Delta _{12} \\ 
\Delta _{21}+\Delta _{22}%
\end{array}%
\right]
\end{equation}%
Since $C1=F,C2=F,C4=F$ we require there\emph{\ no}t to be a solution.

As $|M|=0$ there is \emph{not} a non-zero solution for $\mathbf{v}$ if the
following \emph{requirements} are satisfied%
\begin{equation}
\Delta _{11}\neq 0\quad And/Or\quad \Delta _{12}\neq 0\quad And/Or\quad
\Delta _{21}+\Delta _{22}\neq 0
\end{equation}%
\smallskip

\subsubsection{All Sub-Cases of Case 4}

Overall, we see that if the requirement is satisfied, we can say\textbf{\ }$%
\delta =1$\textbf{.}

Hence applying t\textbf{his result for }$\delta $ in Eq. (\ref%
{Eq.SforNLHVTNew})%
\begin{eqnarray}
S &\leq &\dsum\limits_{\lambda }P(\lambda |c)\left\{
\dsum\limits_{j,k,l,m}P(\alpha _{j},\alpha _{k},\beta _{l},\beta _{m}|\Omega
_{A1},\Omega _{A2},\Omega _{B1},\Omega _{B2};\lambda )\left( 1\right)
\,\right\}  \nonumber \\
&\leq &\dsum\limits_{\lambda }P(\lambda |c)\;(1)  \nonumber \\
&\leq &1
\end{eqnarray}

From above the following \emph{requirements} must be satisfied for Case 4 to
cover all sub-cases.%
\begin{eqnarray}
\Delta _{12} &\neq &0\quad And/Or\quad \Delta _{22}\neq 0\quad And/Or\quad
\Delta _{11}+\Delta _{21}\neq 0\qquad Sub-Case\text{ }a  \nonumber \\
\Delta _{11}+\Delta _{12} &\neq &0\quad And/Or\quad \Delta _{22}\neq 0\quad
And/Or\quad \Delta _{21}\neq 0\qquad Sub-Case\text{ }b  \nonumber \\
\Delta _{11} &\neq &0\quad And/Or\quad \Delta _{22}+\Delta _{12}\neq 0\quad
And/Or\quad \Delta _{21}\neq 0\qquad Sub-Case\text{ }c  \nonumber \\
\Delta _{11} &\neq &0\quad And/Or\quad \Delta _{12}\neq 0\quad And/Or\quad
\Delta _{22}+\Delta _{21}\neq 0\qquad \qquad Sub-Case\text{ }d  \nonumber \\
&&
\end{eqnarray}%
The Bell inequality is then $S\leq 1$. \medskip \pagebreak

\pagebreak


\begin{thebibliography}{99}
\bibitem{Einstein35a} A Einstein, B Podolsky and N Rosen, \textit{Phys. Rev.}
\textbf{47}, 777 (1935).

\bibitem{Schrodinger35a} E Schrodinger, \textit{Naturwissenschaften} \textbf{%
23}, 807 (1935).

\bibitem{Schrodinger35b} E Schrodinger, P\textit{roc. Camb. Phil. Soc.} 
\textbf{31}, 555 (1935).

\bibitem{Bohm52a} D Bohm, \textit{Phys. Rev. }\textbf{85}, 166,180 (1952).

\bibitem{Bell64a} J Bell, \textit{Physics, }\textbf{1}, 195 (1964).

\bibitem{Clauser69a} J Clauser, M Horne, A Shimony and R Holt, \textit{Phys.
Rev. Letts.} \textbf{23}, 880 (1969).

\bibitem{Collins02a} D Collins, N Gisin, N Linden, S Massar and S Popescu, 
\textit{Phys. Rev. Letts.} \textbf{8}8, 040404 (2002).

\bibitem{Dalton21a} B Dalton, \textit{Eur. Phys. J. Spec. Topics} \textbf{230%
}, 903 (2021).

\bibitem{Durr20a} D Durr and D Lazarovici, \textit{Understanding Quantum
Mechanics} (Springer Nature, Switzerland, 2020).

\bibitem{Hiley79a} C Philippidis, C Dewdney and B J Hiley, \textit{Nuov.
Cim. }\textbf{52B}, 15 (1979).

\bibitem{Hall18a} M Ghadini, M Hall and H Wiseman, \emph{ArXiv Quant-Ph }%
1807.01568. (2018).

\bibitem{Everett57a} H Everett, \textit{Rev. Mod.Phys. }\textbf{29}, 454
(1957).

\bibitem{Vaidman21a} L Vaidman, \textit{Many-Worlds Innterpretation of
Quantum Mechanics }Stanford Encyclopedia of Philosopphy (2021).

\bibitem{Leggett03a} A Leggett, \textit{Found. Phys. }\textbf{,33}, 1469
(2003).

\bibitem{Branciard07a} C \ Branciard, A Ling, N Gisin, C Kurtsiefer, A
Lamas-Linares and V Scarani, \textit{Phys.Rev. Letts. }\textbf{99, }210407
(2007).

\bibitem{Wiseman07a} H Wiseman, S Jones and A Doherty, \textit{Phys. Rev.
Letts} \textbf{98}, 140402 (2007).

\bibitem{Reid09a} M D Reid, P D Drummond, W P Bowen, E G Cavalcanti, P K
Lam, H Bachor, U L Anderson and G Leuchs, \textit{Rev. Mod. Phys. }\textbf{81%
}, 1728 (2009).

\bibitem{Aspect82a} A Aspect, J Dalibard and G Roger, \textit{Phys. Rev.
Letts }\textbf{49,} 1804 (1982).

\bibitem{Zeilinger97a} G Weihs, T Jennewein, C Simon, H Weinfurter and A
Zeilinger \textit{Phys. Rev. Letts }\textbf{81}, 5039 (1997).

\bibitem{Werner89a} R\ Werner, \textit{Phys. Rev. A }\textbf{40}, 4277
(1989).

\bibitem{Jones07a} S Jones, H Wiseman, A Doherty, \textit{Phys. Rev. A} 
\textbf{76}, 052116 (2007).

\bibitem{Cavalcanti09a} E Cavalcanti, S Jones, H Wiseman and M Reid, \textit{%
Phys. Rev. A} \ \textbf{80}, 032112 (2009).

\bibitem{Dalton20a} B Dalton, B Garraway and M Reid, \textit{Phys. Rev. A }%
\textbf{101}, 012117 (2020).

\bibitem{Greenberger90a} D Greenberger, M Horne and A Zeilinger, \textit{%
Amer J Phys \ }\textbf{58},1131 (1990).

\bibitem{Leggett80a} A Leggett and A Garg, \textit{Phys. Rev. Letts \ }%
\textbf{54}, 857 (1980).

\bibitem{Teh21a} R Y Teh, L Rosales-Zarate, P D Drummond and M D Reid, 
\textit{Arxiv} 2112.06496 (2021).

\bibitem{Dalton19a} B Dalton, \textit{Eur. Phys. J. Spec. Topics} \textbf{227%
}, 2069 (2019).

\bibitem{Fine82a} A Fine, \textit{Phys. Rev. Lett}s \textbf{48}, 291 (1982)%
{\small .}

\bibitem{Dada11a} A Dada, J.Leach, G Butler, M Padgett and E Anderson, 
\textit{Nat Phys }\textbf{7, }677 (2011).

\bibitem{Dalton14a} B Dalton, L Heaney, J Goold, B Garraway and T Busch, 
\textit{New J Phys} \textbf{16}, 013026 (2014).

\bibitem{Hillary06a} M Hillary and M Zubairy, \textit{Phys. Rev. Letts} 
\textbf{96}, 050503 (2006).
\end{thebibliography}
\end{document}